\documentclass[journal, twocolumn,twoside]{IEEEtran}
\ifCLASSINFOpdf
\else
\fi
\usepackage{url}

\usepackage{amsmath, amsfonts,  amssymb, color, textcomp}
\usepackage[all]{xy}
\usepackage{graphicx}
\usepackage{latexsym}
\usepackage{epsfig}
\usepackage{bm}
\usepackage{cite}
 \usepackage[footnotesize]{caption}
 
 \usepackage{tikz}
\usetikzlibrary{fit,positioning}
\usetikzlibrary{shapes,matrix,decorations.markings,arrows}
\usetikzlibrary{graphs}

\newcommand{\btb}{\mathbf{B}_{tb}^{(\lambda)}}
\newcommand{\bterm}{\mathbf{B}_{[0,L-1]}}
\newcommand{\binf}{\mathbf{B}_{[0,\infty]}}

\renewcommand{\vec}[1]{\ensuremath{\mathbf{#1}}}
\newcommand{\stxt}[1]{\ensuremath{_{\mathrm{#1}}}}

\def\B{\mathbf{B}}
\def\H{\mathbf{H}}
\def\L{L}
\def\l{l}
\def\E{\mathbb{E}}
\renewcommand{\v}[1]{\ensuremath{\mathbf{#1}}}

\newcommand{\epsBP}{\epsilon_{\mathrm{BP}}}
\newcommand{\hBP}{h_\mathrm{BP}(\epsilon)}
\newcommand{\epsMAP}{\epsilon_{\mathrm{MAP}}}

\newcommand{\ie}{\emph{i.e.}}

\newtheorem{example}{Example}
\IEEEoverridecommandlockouts

\definecolor{red}{rgb}{1,0,0}
\usepackage{color,soul}

\definecolor{lightblue}{rgb}{.90,.95,1}
\sethlcolor{yellow}

\usepackage{psfrag}

\usepackage[outdir=./]{epstopdf}

\begin{document}
%
\title{Spatially Coupled Generalized LDPC Codes: Asymptotic Analysis and Finite Length Scaling}

\author{David~G.~M.~Mitchell,~\IEEEmembership{Senior Member,~IEEE,} Pablo M. Olmos,\\Michael~Lentmaier,~\IEEEmembership{Senior Member,~IEEE,} and Daniel~J.~Costello,~Jr.,~\IEEEmembership{Life Fellow,~IEEE,}

        \thanks{This material is based upon work supported in part by the
National Science Foundation under Grant Nos. ECCS-1710920, OIA-1757207 and HRD-1914635, by the European Research Council (ERC) through the European Union’s Horizon 2020 research and innovation program under Grant 714161, and by the Spanish Ministry of Science, Innovation and University under grant TEC2016-78434-C3-3-R (AEI/FEDER, EU). The material in this paper was presented in part at the 2010 IEEE International Symposium on Information Theory, in part at the 2013 IEEE International Symposium on Information Theory, in part at the 2015 IEEE International Symposium on Information Theory, and in part at the 2018 IEEE International Symposium on Turbo Codes \& Iterative Information Processing.}
\thanks{D.~G.~M.~Mitchell is with the Klipsch School of Electrical and Computer Engineering, New Mexico State University, Las Cruces, NM, USA (e-mail: dgmm@nmsu.edu).}
\thanks{P. M. Olmos is with the Signal Theory and Communications Dept., University of Carlos III in Madrid, Legan\'{e}s, Spain (e-mail: olmos@tsc.uc3m.es).}
\thanks{M. Lentmaier is with the Department of Electrical and Information Technology, Lund University, Lund, Sweden (e-mail: Michael.Lentmaier@eit.lth.se).}
\thanks{D.~J.~Costello,~Jr. is with the Department of
Electrical Engineering, University of Notre Dame, Notre Dame, IN, USA
(e-mail: costello.2@nd.edu).}}
\markboth{IEEE Transactions on Information Theory. (Submitted paper)}{}%
\maketitle

\begin{abstract}
Generalized low-density parity-check (GLDPC) codes are a class of LDPC codes in which the standard single parity check (SPC) constraints are replaced by  constraints defined by a linear block code.  These stronger constraints typically result in improved error floor performance, due to better minimum distance and trapping set properties, at a cost of some increased decoding complexity. In this paper, we study spatially coupled generalized low-density parity-check (SC-GLDPC) codes and present a comprehensive analysis of these codes, including: (1) an iterative decoding threshold analysis of SC-GLDPC code ensembles demonstrating capacity approaching thresholds via the threshold saturation effect; (2) an asymptotic analysis of the minimum distance and free distance properties of SC-GLDPC code ensembles, demonstrating that the ensembles are asymptotically good; and (3) an analysis of the finite-length scaling behavior of both GLDPC block codes and SC-GLDPC codes based on a peeling decoder (PD) operating on a binary erasure channel (BEC). Results are compared to GLDPC block codes, and the advantages and disadvantages of SC-GLDPC codes are discussed.
\end{abstract}


\section{Introduction}\label{sec:intro}
Low-density parity-check (LDPC) block codes, with iterative message passing decoding, were introduced by Gallager in 1963 \cite{gal63} as a class of codes whose decoder implementation complexity grows only linearly with block length, in contrast to maximum likelihood (ML) and maximum a posteriori (MAP) decoding methods whose complexity typically has exponential growth.  As a result of the low-density constraint on the parity-check matrix $\mathbf{H}$, the minimum distance of LDPC block codes is sub-optimal.  However, Gallager showed that regular constructions, where the variable and check node degrees of the Tanner graph representation of $\mathbf{H}$ are fixed, maintain linear minimum distance growth with block length, \ie, they are \emph{asymptotically good}, although their iterative decoding thresholds are bounded away from capacity.  Irregular constructions, introduced by Luby et al. in 2001 \cite{lmss01}, where the node degrees are not fixed and can be numerically optimized, have capacity-approaching thresholds, but typically involve a large fraction of {low degree variable nodes that can preclude linear distance growth and result in problematic graphical objects causing failures in sub-optimal decoders}.  As a result, irregular codes perform best in the waterfall, or low signal-to-noise ratio (SNR), portion of the bit-error-rate (BER) performance curve, while regular codes perform better at high SNRs, \ie, in the \emph{error floor} region of the BER curve.

Generalized LDPC (GLDPC) block codes, first proposed by Tanner in 1981 \cite{tan81}, are constructed by replacing some/all of the single parity-check (SPC) constraint nodes in the Tanner graph of a conventional LDPC code by more powerful generalized constraint (GC) nodes corresponding to an $(n,k)$ linear block code.  The $n$ variable nodes connected to a GC node in the Tanner graph of a GLDPC code are then considered as the code bits of the corresponding $(n,k)$ code, and the sub-code associated with each GC node is referred to as a \emph{constraint code}.  In message passing decoding of GLDPC codes, the constraint codes are decoded using standard block code decoders which, in the case of simple constraint codes such as Hamming codes \cite{lz99} or Hadamard codes \cite{ypw07}, can be ML or MAP decoders.  GLDPC codes have many potential advantages compared to conventional SPC/LDPC codes, such as large minimum distance \cite{lz99,bpz99}, low error floors \cite{lrc08}, and fast iterative decoding convergence \cite{mpf15}.

Spatially coupled LDPC (SC-LDPC) codes, also known as LDPC convolutional codes, were introduced by Jimenez-Felstrom and Zigangirov in 1999 \cite{fz99}.  SC-LDPC codes can be viewed as a sequence of LDPC block codes whose graph representations are coupled together over time, resulting in a convolutional structure with block-to-block memory.  A remarkable property of SC-LDPC codes, established numerically in \cite{lscz10} and analytically in \cite{kru11,kru13}, is that {asymptotically} their iterative message passing decoding threshold is equal to the MAP decoding threshold of the underlying LDPC block code ensemble {under certain conditions}, a phenomenon known as \emph{threshold saturation}.  In other words, the (exponential complexity) MAP decoding performance of the underlying block code can be achieved by its coupled version with (linear complexity) message passing decoding. 

{Spatially coupled LDPC codes with generalized constraints, or \emph{spatially coupled generalized LDPC (SC-GLDPC) codes}, and related constructions, including braided codes} \cite{ftlz09,zlzc10}{, staircase codes} \cite{sfh12}{, and product codes }\cite{jpn+13}{, have been investigated in the literature} \cite{lf10,jian13,mlc13,omc15,jpn17,hpgb17,ztk18,cmol18}{. In particular, it has been shown that SC-GLDPC codes have good iterative decoding thresholds }\cite{lf10,hpgb17}{, including excellent performance with hard decision iterative decoding }\cite{jian13, jpn17, ztk18}{, linear growth of minimum distance }\cite{mlc13}{, and robust finite-length scaling performance }\cite{cmol18}{. We note that most of the existing work in this area can be considered as spatially coupled versions of product codes. Staircase codes, for example, can be seen as a variation of tightly braided block codes where the graphs are deterministic and less sparse. In particular, unlike the GLDPC code ensembles we consider in this paper, such codes get sparser only when the length of the component codes is increased.} 

Motivated to combine the threshold improvement of spatial coupling with the improved distance properties of generalized constraints, this {paper investigates SC-GLDPC codes with linear block codes as generalized constraints} and presents both asymptotic (threshold and distance) and finite-length analyses of SC-GLDPC code ensembles.  A principle contribution of this paper is to extend the results of \cite{lf10,mlc13,omc15,cmol18} and present a unified treatment and analysis of SC-GLDPC codes. We first extend the threshold analysis of protograph-based SC-LDPC code ensembles in \cite{lscz10} to GLDPC code ensembles and use this to perform an iterative decoding threshold analysis of SC-GLDPC codes ensembles. This method is used to show numerically that threshold saturation is achieved for SC-GLDPC code ensembles, \ie, their thresholds coincide with the maximum a-posteriori (MAP) decoding threshold of the underlying GLDPC block code ensemble.\footnote{In a recent paper \cite{yap18}, the authors found that, for a particular class of \emph{doubly-generalized} LDPC codes introduced in \cite{wf06}, in which both variable and check nodes have generalized constraints, no threshold improvement from spatial coupling is observed.  However, as we will demonstrate in this paper, threshold improvement is achieved with SC-GLDPC codes.} This is followed by a minimum distance analysis of terminated and tail-biting SC-GLDPC code ensembles and a free distance analysis of unterminated ensembles, both of which demonstrate that the ensembles are asymptotically good and have large distance growth rates.  In order to study the finite-length scaling properties of SC-GLDPC code ensembles, a method to analyze the finite-length scaling behavior of GLDPC block codes over the binary erasure channel (BEC) with peeling decoding (PD) is first introduced. We then extend this approach to study SC-GLDPC code ensembles and demonstrate robust finite-length scaling performance.  

\section{Protograph-Based SC-GLDPC Codes}\label{sec:proto}
A protograph \cite{tho03} is a small bipartite graph that connects a set of $n_v$ variable nodes $V=\{v_1,v_2,\ldots,v_{n_v}\}$ to a set of $n_c$  constraint nodes $C=\{c_1,c_2,\ldots,c_{n_c}\}$ by a set of edges $E$. The edges connected to a variable node $v_j$ of degree $\partial(v_j)$ or a constraint node $c_i$ of degree $\partial(c_i)$ are labeled by $e^\mathrm{v}_{j,a}$ or $e^\mathrm{c}_{i,b}$, respectively, where {$a\in\{1,\dots,\partial(v_j)\}$} and {$b\in\{1,\dots,\partial(c_i)\}$}. If the $a$-th edge associated with $v_j$ is the $b$-th edge associated with $c_i$, then $e^\mathrm{v}_{j,a}=e^\mathrm{c}_{i,b}$.\footnote{This way of labeling will be useful for the density evolution equations described in Section~\ref{sec:thres}, since it takes into account  the order of edges connected to a node and allows one to distinguish among multiple edges between a given pair of nodes.} 
In a protograph-based GLDPC code ensemble, each constraint node $c_i$ can represent an arbitrary block constraint code $\mathcal{C}_i$ with parity-check matrix $\mathbf{H}_{c_i}$, length $n^{c_i}$, and $m^{c_i}$ linearly independent parity-check equations {where, throughout the manuscript, we use superscript labels on code parameters in order to distinguish between different constraint codes}. The \emph{design rate} of the GLDPC code ensemble is then given by
\begin{equation}\label{blockrate}
R = 1 - \frac{\sum_{i=1}^{n_c}m^{c_i}}{n_v}.
\end{equation}

A protograph can be represented by means of an $n_c\times n_v$ bi-adjacency matrix $\mathbf{B}$, which is called the \emph{base matrix} of the protograph. The nonnegative integer entry $B_{ij}$ in row $i$ and column $j$ of $\mathbf{B}$ is equal to the number of edges that connect nodes $c_i$ and $v_j$ in the protograph. In order to construct ensembles of protograph-based GLDPC codes, a protograph can be interpreted as a template for the Tanner graph of a derived code, which is then obtained by a \emph{copy-and-permute} or \emph{graph lifting} operation \cite{tho03}. In matrix form, the protograph is lifted by replacing each nonzero entry $B_{ij}$ of $\mathbf{B}$ with a summation of $B_{ij}$ non-overlapping permutation matrices of size $M\times M$, thereby creating an $Mn_c \times Mn_v$ constraint matrix $\mathbf{H}$ of a GLDPC code. Each row in the $i^\text{th}$ set of $M$ rows of $\mathbf{H}$ must satisfy the constraints $\mathbf{H}_{c_i}$ associated with constraint node $c_i$, where the length $n^{c_i}$ of the $i^\text{th}$ constraint code equals the sum of the entries in the $i^\text{th}$ row of $\mathbf{B}$ and the constraint applies to the positions in a row of $\mathbf{H}$ with non-zero entries.\footnote{Strictly speaking, $\mathbf{H}$ is not a parity-check matrix since each row in the $i^\text{th}$ set of $M$ rows of $\mathbf{H}$ corresponds to $m^{c_i}$ parity-checks. Consequently, we refer to $\mathbf{H}$ as a constraint matrix.} Allowing the permutations to vary over all $M!$ possible choices results in an ensemble of GLDPC block codes. 

\begin{figure}[t]
\begin{center}
\includegraphics[width=2.0in]{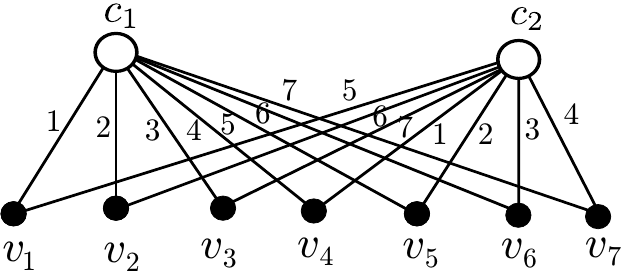}
\end{center}
\caption{Protograph of a $(2,7)$-regular GLDPC block code. The white circles represent generalized constraint nodes and the black circles represent variable nodes. The labels on the edges indicate the corresponding columns of the {parity-check} matrix $\mathbf{H}_{c}$ of the generalized constraint code. In this case, both constraints are defined by the same $(7,4)$ Hamming code, but with {different orderings of columns}.}\label{fig:27ham}
\end{figure}
\begin{figure*}[t]
\begin{center}
\includegraphics[width=5.1in]{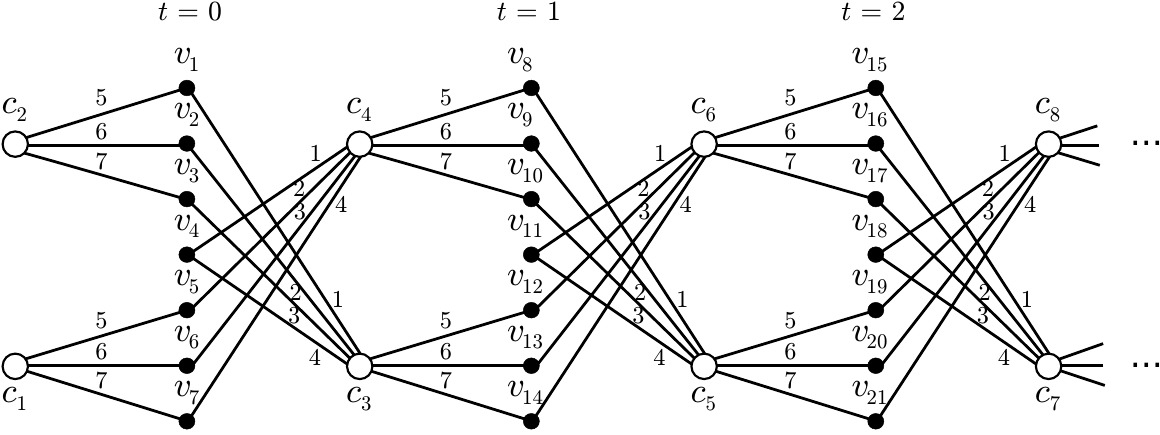}
\end{center}
\caption{Convolutional protograph of the $(2,7)$-regular SC-GLDPC code ensemble $A_7$. The white circles represent generalized constraint nodes and the black circles represent variable nodes.}\label{fig:scprot}
\end{figure*}
\begin{example}\label{ex:blockhamming} Fig.~\ref{fig:27ham} displays the protograph of an  $(n_c,n_v)=(2,7)$-regular GLDPC block code with base matrix
\begin{equation}
\mathbf{B}=\left[
\begin{array}{ccccccc}
1 & 1 & 1 & 1 & 1 & 1 & 1\\
1 & 1 & 1 & 1 & 1 & 1 & 1
\end{array}\right].\label{hammingbase}
\end{equation}
If we suppose both the constraint nodes are $(7,4)$ Hamming codes with parity-check matrix
\begin{equation}\label{hammingmatrices}\mathbf{H}_c=\left[\begin{array}{ccccccc}
1 & 0 & 0 & 1 & 1 & 1 & 0\\
0 & 1 & 0 & 1 & 1 & 0 & 1\\
0 & 0 & 1 & 1 & 0 & 1 & 1
\end{array}\right], 
\end{equation}
where the constraint code length is $n^c=7$ and the row rank of $\mathbf{H}_c$ is $m^c=3$, then the resulting  ensemble has design rate $R=1/7$. Note that even though both constraints are defined by the same $(7,4)$ Hamming code, a different ordering of columns can be used. In Fig.~\ref{fig:27ham}, the column of $\mathbf{H}_c$ that the variable node is connected to is shown on the edge.\hfill $\Box$
\end{example}

\subsection{Unterminated SC-GLDPC codes}\label{sec:convproto}
An unterminated SC-GLDPC code can be described by a {\em convolutional protograph} \cite{mlc15} with base matrix
\begin{equation}\label{convbase}\mathbf{B}_{[0,\infty]}=\left[
\begin{array}{cccccc}
\mathbf{B}_{0} &   & \\
\mathbf{B}_{1} &\mathbf{B}_{0} &  \vspace{-2.2mm}\\
\vdots &\mathbf{B}_{1} & \vspace{-2mm}\ddots \\
\mathbf{B}_{w} & \vdots& \ddots\\
 & \mathbf{B}_{w}& \vspace{-2mm}\\
 & & \ddots\\
\end{array}\right],\end{equation}
where $w$ denotes the \emph{syndrome former memory} or \emph{coupling width} of the code and the $b_c \times b_v$ {\em component base matrices} $\mathbf{B}_{i}$, {$i\in\{0,1,\ldots,w\}$}, represent the edge connections from the $b_v$ variable nodes at time $t$ to the $b_c$ (generalized) constraint nodes at time $t+i$.\footnote{{We note that in a series of papers by Kudekar et al.} \cite{kru11,kru13},{ the authors refer to a ``smoothing parameter'' $w$ which is equal to the syndrome former memory plus one.}} An ensemble of (in general) time-varying SC-GLDPC codes can then be formed from $\mathbf{B}_{[0,\infty]}$ using the protograph construction method  described above with lifting factor $M$.\footnote{{In this paper, we restrict our attention to time-invariant protographs. Consequently, the choice of permutations in the graph lifting stage will determine if the lifted graph is time-invariant or time-varying. A random lifting will typically result in a time-varying graph.}} The \emph{decoding constraint length} of the resulting ensemble is given by $\nu_s = (w + 1)Mb_v$, the \emph{design rate} is given by
\begin{equation}\label{untermrate}
R = 1 - \frac{\sum_{i=1}^{b_c}m^{c_i}}{b_v},
\end{equation}
and at each time instant $t$ the encoder creates a block $\mathbf{v}_t$ of $Mb_v$ symbols resulting in the unterminated code sequence $\mathbf{v}=[\mathbf{v}_0,\mathbf{v}_1,\ldots,\mathbf{v}_t,\ldots$].

Starting from a $b_c \times b_v$ base matrix $\mathbf{B}$ of a block code ensemble, one can construct SC-GLDPC code ensembles with the same variable and check node degrees as $\mathbf{B}$. This is achieved by an {\em edge spreading} procedure \cite{mlc15}  that divides the edges connected to each variable node in the base matrix $\mathbf{B}$ among $w+1$ component base matrices $\mathbf{B}_i$, {$i\in\{0,1,\ldots,w\}$}, such that the condition $\mathbf{B}_0+\mathbf{B}_1+\cdots+\mathbf{B}_{w}=\mathbf{B}$ is satisfied. We now give some examples of constructing SC-GLDPC code ensembles.

\begin{example} \label{ex:convhamming}For $w=1$, we can apply the edge spreading technique to the $(b_c, b_v) = (2,7)$-regular block code base matrix in \eqref{hammingbase} to obtain the following component base matrices
\begin{eqnarray}
&\mathbf{B}_0=\left[
\begin{array}{ccccccc}
0 & 0 & 0 & 0 & 1 & 1 & 1\\
1 & 1 & 1 & 0 & 0 & 0 & 0
\end{array}\right],\label{b1}\\
&\mathbf{B}_1=\left[
\begin{array}{ccccccc}
1 & 1 & 1 & 1 & 0 & 0 & 0\\
0 & 0 & 0 & 1 & 1 & 1 & 1
\end{array}\right].\label{b2}
\end{eqnarray}
The convolutional protograph associated with the resulting base matrix $\mathbf{B}_{[0,\infty]}$ defined in \eqref{convbase} is shown in Fig. \ref{fig:scprot}, where time indices $t$ are shown above the corresponding set (block) of variable nodes. We choose the upper and lower constraint nodes at each time instant to correspond to the $(7,4)$ Hamming code with $n^c=7$, $m^c=3$, and {parity-check} matrix $\mathbf{H}_c$ from \eqref{hammingmatrices}. (Note that the labels indicated on the edges correspond to columns of the component parity-check matrix $\mathbf{H}_c$  in \eqref{hammingmatrices} and that the constraint nodes $c_1$ and $c_2$ represent shortened codes.) After lifting, the constraint length of the resulting SC-GLDPC code ensemble is $\nu_s = 14M$ and the design rate is $R=1/7$.

We will refer to this SC-GLDPC code ensemble as Ensemble $A_7$. 
An extension to the Ensemble $A_{15}$ representing design rate $R=7/15$ SC-GLDPC codes corresponding to the $(b_c,b_v)=(2,15)$-regular all-ones matrix $\mathbf{B}$, $(15,11)$ Hamming constraint codes with $n^c=15$ and $m^c=4$, and $w=1$ edge-spreading based on \eqref{b1} and \eqref{b2}, and to other values of $b_v=n^c$, $m^c$, and $R$, is straightforward.\hfill $\Box$\end{example}

To illustrate the flexibility of SC-GLDPC code designs, multiple edges can also be introduced in the block protograph.

\begin{figure}
\begin{center}
\includegraphics[width=3in]{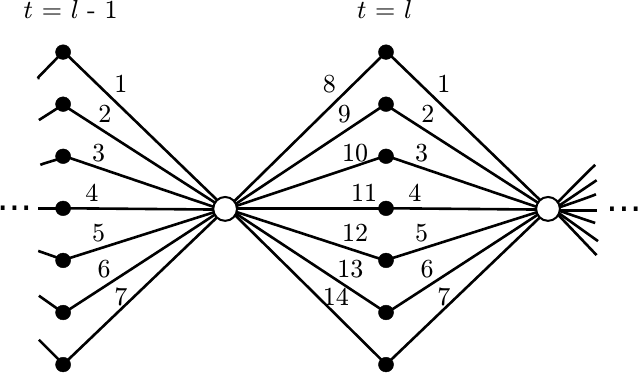}
\caption{Segment of the $w=1$ convolutional protograph defining Ensemble $B_{14}$.} \label{fig:PG7_14} \vspace{-1em}
\end{center}
\end{figure}

\begin{example}  \label{bcc14} Considering shortened $(14,10)$ Hamming constraint codes with $n^c=14$ and $m^c=4$ as an example, where each variable node in a $b_c\times b_v = 1 \times 7$ protograph is connected with a double edge to a single check node. We split the  corresponding multi-edge base matrix $\vec{B}=[\,\begin{matrix}
2 & 2 & 2 & 2 & 2 & 2 & 2 \\
\end{matrix}\,]$ into
\begin{equation}
\vec{B}_0=\vec{B}_1=\begin{bmatrix}
1 & 1 & 1 & 1 & 1 & 1 & 1 \\
\end{bmatrix}
\end{equation}
and obtain the protograph of Ensemble $B_{14}$ with $w=1$, $b_c=1$ check node, and $b_v=7$ variable nodes at each time instant, a segment of which is illustrated in Fig.~\ref{fig:PG7_14}. From \eqref{untermrate} we see that the design rate  of Ensemble $B_{14}$ is $R=1-4/7 = 3/7$. Puncturing the first variable node at each time instant $t$  results in Ensemble $B_{14,P}$ with design rate $R=0.5$. \hfill $\Box$\end{example}

The protograph-based \emph{braided block code} (BBC) ensembles considered in \cite{lnf10} are another example of SC-GLDPC code ensembles. These can be derived by using the Tanner graph of a \emph{tightly} BBC \cite{ftlz09} as a protograph. The component base matrices of such an SC-GLDPC code can be identified as
 \begin{equation}\label{bbc}
\vec{B}_0=\begin{bmatrix}
1 & \vec{i} & \vec{0} \\
1 & \vec{0} & \vec{i}
\end{bmatrix}, \ 
\vec{B}_i=\begin{bmatrix}
0 & \vec{0} & \vec{e}_i \\
0 & \vec{e}_i & \vec{0}
\end{bmatrix}, 
\end{equation}
where  $i=1,\dots, w$, $\vec{e}_i=(0,\dots,0,1,0,\dots,0)$ is the length $w$ vector with a one at the $i^\text{th}$ position and zeros elsewhere, $\vec{0}$ is the all-zero vector, and  $\vec{i}$ the all-one vector, of length $w$. We will use the term Ensemble $C_{n^c}$ when referring to such SC-GLDPC code ensembles based on constraint codes of length $n^c=2w+1$.  

\begin{example} \label{ex:bcc7} For the $(b_c,b_v)=(2,7)$-regular GLDPC base matrix with $(7,4)$ Hamming constraint codes and design rate $R=1/7$ from Example \ref{ex:blockhamming}, the convolutional protograph resulting from the tightly BBC construction of \eqref{bbc} with $w=3$, $n^c=7$, and $m^c=3$, corresponding to Ensemble $C_7$, is illustrated in Fig.~\ref{fig:TG_Tight27}, where the upper constraint nodes correspond to the ``horizontal constraints" and the lower constraint nodes correspond to the ``vertical constraints" of the braided construction. Its girth is equal to eight, which follows from the structure of the array and is true for any SC-GLDPC code resulting from a tightly BBC protograph.
\begin{figure*}
\begin{center}
\includegraphics[width=6.1in]{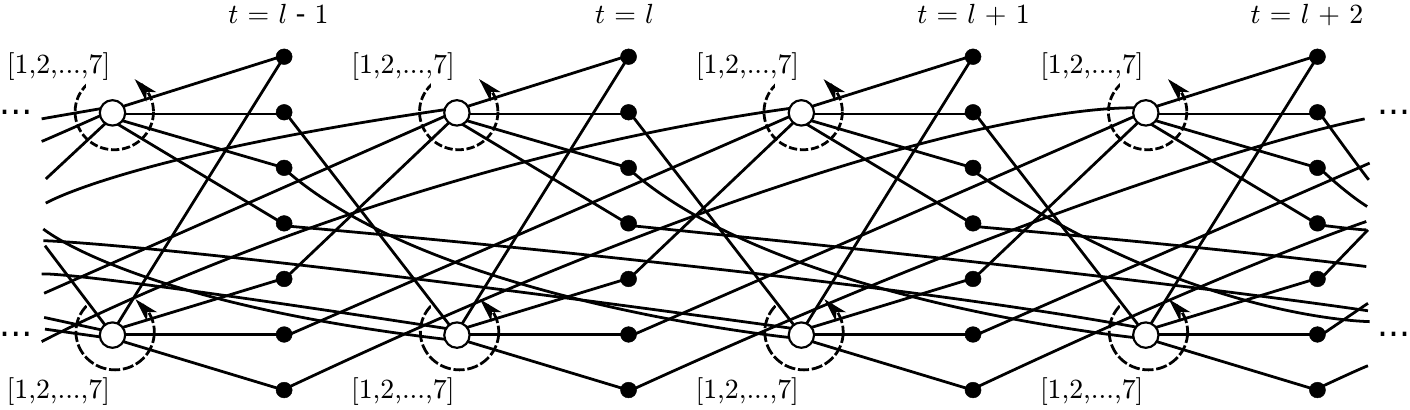}
\caption{Convolutional protograph of a SC-GLDPC code with $(7,4)$ Hamming constraint codes, defining Ensemble $C_7$. The nodes are grouped according to the time instant $t$ at which the code symbols, designated by the filled circles, are generated. Edge labels $1,2,\ldots,7$, corresponding to the columns of the component matrix $\v{H}_c$, are ordered anti-clockwise from the upper left of each constraint node.} \label{fig:TG_Tight27} 
\end{center}
\end{figure*}

Observe that the sum of the component base matrices  in \eqref{bbc} is equal to the base matrix $\mathbf{B}$ in \eqref{hammingbase} of the corresponding GLDPC code, \ie, the all-one matrix of dimension $b_c \times b_v = 2 \times 7$. This reflects the fact that the graph in Fig.~\ref{fig:TG_Tight27} can be obtained by repeating the GLDPC graph in Fig.~\ref{fig:27ham} and spreading the edges over $w=3$ adjacent time instants. An extension to the Ensemble $C_{15}$, representing SC-GLDPC codes corresponding to the $(b_c,b_v)=(2,15)$-regular all-one base matrix $\mathbf{B}$, $(15,11)$ Hamming constraint codes, and $w=7$ edge-spreading based on \eqref{bbc} with $n^c=15$, $m^c=4$, and design rate $R=7/15$, as well as to other values of $b_v=n^c$, $m^c$, and $R$, is straightforward.\hfill $\Box$\end{example}

\subsection{Terminated and tail-biting SC-GLDPC codes}\label{sec:term}
\begin{figure*}[t]
\begin{center}
\includegraphics[width=5.2in]{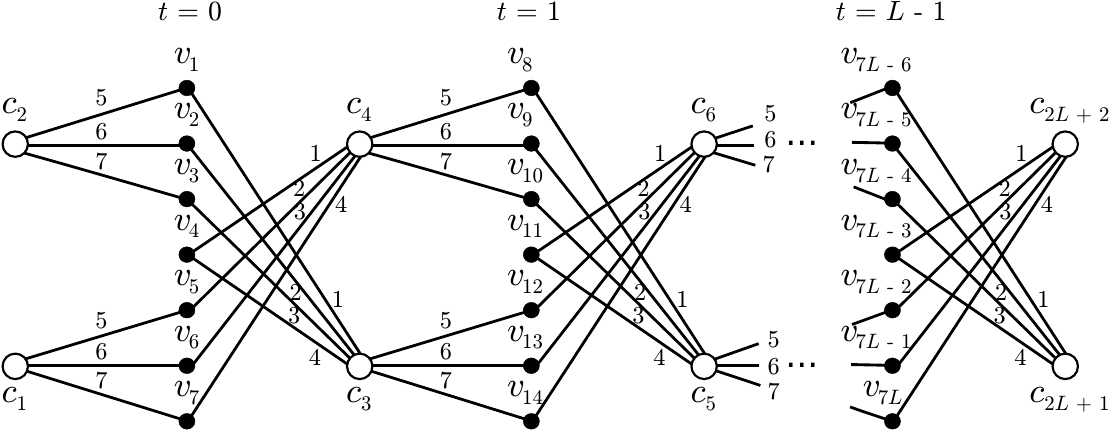}
\end{center}
\caption{Protograph of the $(2,7)$-regular terminated SC-GLDPC code ensemble $A_7$ with coupling length $L$.}\label{fig:scprotterm}
\end{figure*}

Suppose that we start the SC-GLDPC code with convolutional base matrix defined in $(\ref{convbase})$ at time $t=0$ and terminate it after $L$ time instants. The resulting finite-length base matrix is then given by 
\begin{equation}\label{termbase}\mathbf{B}_{[0,L-1]}=\left[
\begin{array}{cccc}
\mathbf{B}_0 & & &\\
\mathbf{B}_1 &\mathbf{B}_0 & &\\
\vdots &\mathbf{B}_1 & \ddots &\\
\mathbf{B}_{w} &\vdots & \ddots & \mathbf{B}_0 \\
&\mathbf{B}_{w} &  & \mathbf{B}_1 \\
&& \ddots & \vdots\\
&& & \mathbf{B}_{w}
\end{array}\right]_{(L+w)b_c \times Lb_v},
\end{equation}
where $L$ is called the \emph{coupling length}. The matrix $\mathbf{B}_{[0,L-1]}$ is then the base matrix of a \emph{terminated SC-GLDPC code}. The corresponding terminated convolutional protograph is slightly irregular, with lower constraint node degrees at both ends. This is illustrated for the $A_7$ ensemble of Example~\ref{ex:convhamming} in Fig.~\ref{fig:scprotterm}. The reduced degree constraint nodes at each end of the convolutional protograph are associated with shortened constraint codes, in which the symbols corresponding to the missing edges are removed. For decoding purposes, such a code shortening is equivalent to fixing these removed symbols and assigning an infinite reliability to them. Note that the variable node degrees are not affected by termination.

The constraint matrix $\mathbf{H}_{[0,L-1]}$ of the terminated SC-GLDPC code derived from $\mathbf{B}_{[0,L-1]}$ by lifting with some factor $M$  has $Mb_v L$ columns and $(L+w)Mb_c$ rows. It follows that the rate of the terminated SC-GLDPC code is equal to
\begin{equation}\label{convrate}R_L =1-\frac{(L+w)b_cm^c-\Delta}{Lb_v},\end{equation}
where $m^c$ denotes the (constant) number of independent parity checks associated with each constraint code and $\Delta\geq 0$ accounts for a possible rate increase due to the shortened constraint codes.\footnote{We assume here the simplified case where each generalized constraint code is described by a parity-check matrix $\v{H}_c$ with $m^c$ independent parity checks. Under this assumption, the rate formula in \eqref{blockrate} for GLDPC block codes becomes $R=1-b_cm^c/b_v$.} If $\mathbf{H}_{[0,L-1]}$ has full rank, the rate increase parameter is $\Delta = 0$. However, the shortened constraint codes at the ends of the graph can cause a reduced rank for $\mathbf{H}_{[0,L-1]}$, which slightly increases $R_L$. In this case,  $\Delta>0$ and depends on both the particular constraint code chosen and the assignment of edges to the columns of its parity-check matrix $\mathbf{H}_{c}$. As $L\rightarrow\infty$, the rate $R_L$ of the terminated SC-GLDPC code converges to the design rate $R=1-b_cm^c/b_v$ of the underlying GLDPC block code with base matrix $\mathbf{B}$.\footnote{We note here that the $(L+w)Mb_c$ rows of $\mathbf{H}_{[0,L-1]}$ should be viewed as $(L+w)b_c$ groups of rows, with $M$ entries in each group, that are decoded according to the same constraint code with $m^c$ rows.}

The generalized convolutional base matrix $\mathbf{B}_{[0,\infty]}$ can also be terminated using \emph{tail-biting} \cite{st79,mw86}, resulting in the base matrix of a tail-biting generalized LDPC (TB-GLDPC) code ensemble. Here, for any $\lambda \geq w$, the last $b_c w$ rows of the terminated parity-check matrix $\mathbf{B}_{[0,\lambda-1]}$ are removed and added to the first $b_c w$ rows to form the $\lambda b_c \times \lambda b_v$ tail-biting parity-check matrix $\mathbf{B}_{tb}^{(\lambda)}$ with tail-biting coupling length $\lambda$ 
\begin{equation}\label{basetb}\mathbf{B}_{tb}^{(\lambda)} =
\scalebox{0.9}{\mbox{$\left[\begin{array}{cccccccccc}
\mathbf{B}_0 &  &&& &   &  & \mathbf{B}_{w}   & \cdots &  \mathbf{B}_1 \\
\mathbf{B}_1 & \mathbf{B}_0  & &&&  &    &  & \ddots& \vdots \\
\vdots & \vdots  &&&&   &  &  &    & \mathbf{B}_{w}  \\
\mathbf{B}_{w} & \mathbf{B}_{w-1}  & &&&  &  &  &    & \\
 & \mathbf{B}_{w}  & &&& \ddots  &  &  &  &  \\
 &   & &&&  & \mathbf{B}_{0} &  &  &    \\
 &   &&&& \ddots  & \vdots & \mathbf{B}_{0}  &  &  \\
 &   & &&& \ddots & \mathbf{B}_{w-1}  & \vdots & \ddots &    \\
 &  & &&&& \mathbf{B}_{w} & \mathbf{B}_{w-1} & \cdots &\mathbf{B}_{0}   \\
\end{array}\right].$}}\end{equation}
\noindent  Note that, if $w=1$ and $\lambda = 1$, the tail-biting base matrix is simply the original block code base matrix, \ie, $\mathbf{B}_{tb}^{(1)}=\mathbf{B}$. 
Terminating $\mathbf{B}_{[0,\infty]}$ in such a way preserves the design rate of the ensemble, \ie, $R_\lambda =1-\lambda b_c m^c/\lambda b_v=1-b_cm^c/b_v=R$, and we see that $\mathbf{B}_{tb}^{(\lambda)}$ has exactly the same degree distribution as the original block code base matrix $\mathbf{B}$.

\section{Threshold Analysis of SC-GLDPC Codes}\label{sec:thres}
We assume belief propagation (BP) decoding, with log-likelihood ratios (LLRs) acting as messages. At every iteration $\ell$, first all constraint nodes and then all variable nodes are updated. The outgoing messages computed at a constraint node $c_i$ {toward its $b^{\text{th}}$ neighboring variable node} at iteration $\ell$ are then equal to
\begin{align}
L_\mathrm{c}^{(\ell)}(e_{i,b}^\mathrm{c})  = \log \sum_{\vec{x} \in \mathcal{C}_i^{b,0}} &\prod_{b^\prime \neq b} \exp\left( L_\mathrm{v}^{(\ell-1)}(e_{i,b^\prime}^\mathrm{c}) (1/2 - x_{b^\prime})\right)  \notag \\
 - \log \sum_{\vec{x} \in \mathcal{C}_i^{b,1}}  \prod_{b^\prime \neq b} \exp & \left(  L_\mathrm{v}^{(\ell-1)}(e_{i,b^\prime}^\mathrm{c}) (1/2 - x_{b^\prime})\right), \label{eq:BPCheckUpdateGLDPC}
\end{align}
where $b,b^\prime \in \{1,\dots,\partial(c_i)\}$, $L_\vec{v}^{(\ell-1)}(e^\mathrm{c}_{i,b^\prime})$ is the LLR received at constraint node $c_i$ from the variable node connected to $e^\mathrm{c}_{i,b^\prime}$ at iteration $\ell$, and we have partitioned $\mathcal{C}_i$ into the sets $\mathcal{C}_i^{b,0}$ and $\mathcal{C}_i^{b,1}$, corresponding to codewords $\vec{x}=[\begin{array}{cccc} x_1 & x_2 & \cdots &x_{\partial(c_i)}\end{array}]\in\mathcal{C}_i$ for which $x_b=0$ and $x_b=1$, respectively. The message $L_\mathrm{c}^{(\ell)}(e_{i,b}^\mathrm{c})$ corresponds to the $b$-th extrinsic output generated by an optimal \emph{a posteriori probability} (APP) decoder for component code $\mathcal{C}_i$, which is computed from the incoming messages $L_\mathrm{v}^{(\ell-1)}(e^\mathrm{c}_{i,b^\prime})$, $b^\prime\neq b$, to constraint node $c_i$. The incoming messages of the first iteration are initialized by the channel LLRs $L_\mathrm{ch}(v_j)$ of the neighboring variable nodes, \ie,  $L_\mathrm{v}^{(0)}(e^\mathrm{c}_{i,b^\prime})=L_\mathrm{ch}(v_j)$, where $v_j$ is the variable node connected to $e^\mathrm{c}_{i,b^\prime}$. The outgoing messages computed at a variable node $v_j$ at iteration $\ell$ are equal to
\begin{equation} \label{eq:BPVariableUpdate}
L_\mathrm{v}^{(\ell)}(e_{j,a}^\mathrm{v}) = L_\mathrm{ch}(v_j) + \sum_{a^\prime \neq a} L_\mathrm{c}^{(\ell)}(e_{j,a^\prime}^\mathrm{v}), 
\end{equation}
where $a,a^\prime \in \{1,\dots,\partial(v_j)\}$. 

\subsection{Density evolution for GLDPC code ensembles}\label{sec:thressub}
For transmission over a  binary erasure channel (BEC), the messages that are passed between the nodes represent either an erasure or the correct symbol values $0$ or $1$. {In this case, the} BP decoder is particularly simple and exact density evolution can be described explicitly.  
Let $q^{(\ell)}(e^\mathrm{c}_{i,b})$ denote the probability that the check to variable node message sent along edge $e^\mathrm{c}_{i,b}$ in decoding iteration $\ell$ is an erasure. 
Assuming a conventional LDPC code ensemble, where $c_i$ corresponds to an SPC code, this is the case if at least one of the incoming messages from the other neighboring variable nodes is erased, \ie,
\begin{equation}\label{eq:CheckUpdateLDPC}
q^{(\ell)}(e^\mathrm{c}_{i,b}) = 1 - \prod_{b^\prime \neq b} \left( 1 - p^{(\ell-1)}(e^\mathrm{c}_{i,b^\prime}) \right),
\end{equation}
where the $p^{(\ell-1)}(e^\mathrm{c}_{i,b^\prime})$, $b,b^\prime \in \{1,\ldots,\partial(c_i)\}$, denote the probabilities that the incoming messages to $c_i$ computed in the previous iteration are erasures. For a GLDPC code ensemble, where $c_i$ corresponds to an arbitrary block code, \eqref{eq:CheckUpdateLDPC} can be replaced by the general expression  
\begin{equation}\label{eq:CheckUpdateGLDPC}
q^{(\ell)}(e^\mathrm{c}_{i,b}) = f^{\mathcal{C}_i}_b\left(p^{(\ell-1)}(e^\mathrm{c}_{i,b^\prime}), b^\prime \neq b \right),
\end{equation}
where $f^{\mathcal{C}_i}_b$ is a multi-dimensional input/output transfer function that characterizes the APP decoder that computes the messages $L_\mathrm{c}^{(\ell)}(e_{i,b}^\mathrm{c})$ corresponding to \eqref{eq:BPCheckUpdateGLDPC}. Note that, for generalized codes,
$f^{\mathcal{C}_i}_b$ can be different for each $b \in \{1,\dots,\partial(c_i)\}$, which implies that the ordering of edges can affect the performance of the ensemble.
A method for computing explicit expressions for the  APP decoder output distributions
that can be used in (\ref{eq:CheckUpdateGLDPC}) was presented in \cite{ltf09}. This method is based on a Markov chain analysis of the decoder metrics using a trellis representation of the block code $\mathcal{C}_i$.

The variable to check node message sent along edge $e^\mathrm{v}_{j,a}$ is an erasure if all incoming messages from the channel and from the other neighboring check nodes are erasures. Thus we have
\begin{equation}\label{eq:VariableUpdate}
p^{(\ell)}(e^\mathrm{v}_{j,a}) = \epsilon \prod_{a^\prime \neq a} q^{(\ell)}(e^\mathrm{v}_{j,a^\prime}),
\end{equation}
where $a,a^\prime \in \{1,\dots,\partial(v_j)\}$ and $\epsilon$ is the erasure probability of the BEC. The largest channel value $\epsilon$ for which \eqref{eq:CheckUpdateGLDPC} and \eqref{eq:VariableUpdate} converge, denoted $\epsBP$, is the \emph{threshold} of the BP decoder for the GLDPC code ensemble.

\subsection{{Bounding MAP thresholds with BP extrinsic information transfer functions}}
The extrinsic probability $p_\mathrm{BP,extr}(v_j,\epsilon)$ that a symbol associated with variable node $v_j$ remains erased after $\ell$ iterations of BP decoding can be expressed as
\begin{equation}
p\stxt{BP,extr}(v_j,\epsilon) = \prod_{a} q^{(\ell)}(e^\mathrm{v}_{j,a}).
\end{equation}
Note that here the product is  over {\em all} incoming messages to $v_j$ and the channel erasure probability $\epsilon$ does not appear in the expression but implicitly involved in the calculation of $q^{(\ell)}(e^\mathrm{v}_{j,a})$. The BP extrinsic information transfer (EXIT) function $\hBP$ \cite{mmu08} is given by the average of $p_\mathrm{BP,extr}(v_j,\epsilon)$ over all transmitted variable nodes {$v_j\in V$, \ie, the average is computed excluding all of the punctured variable nodes.}

\begin{figure}
\begin{center}
\includegraphics[width=3.5in]{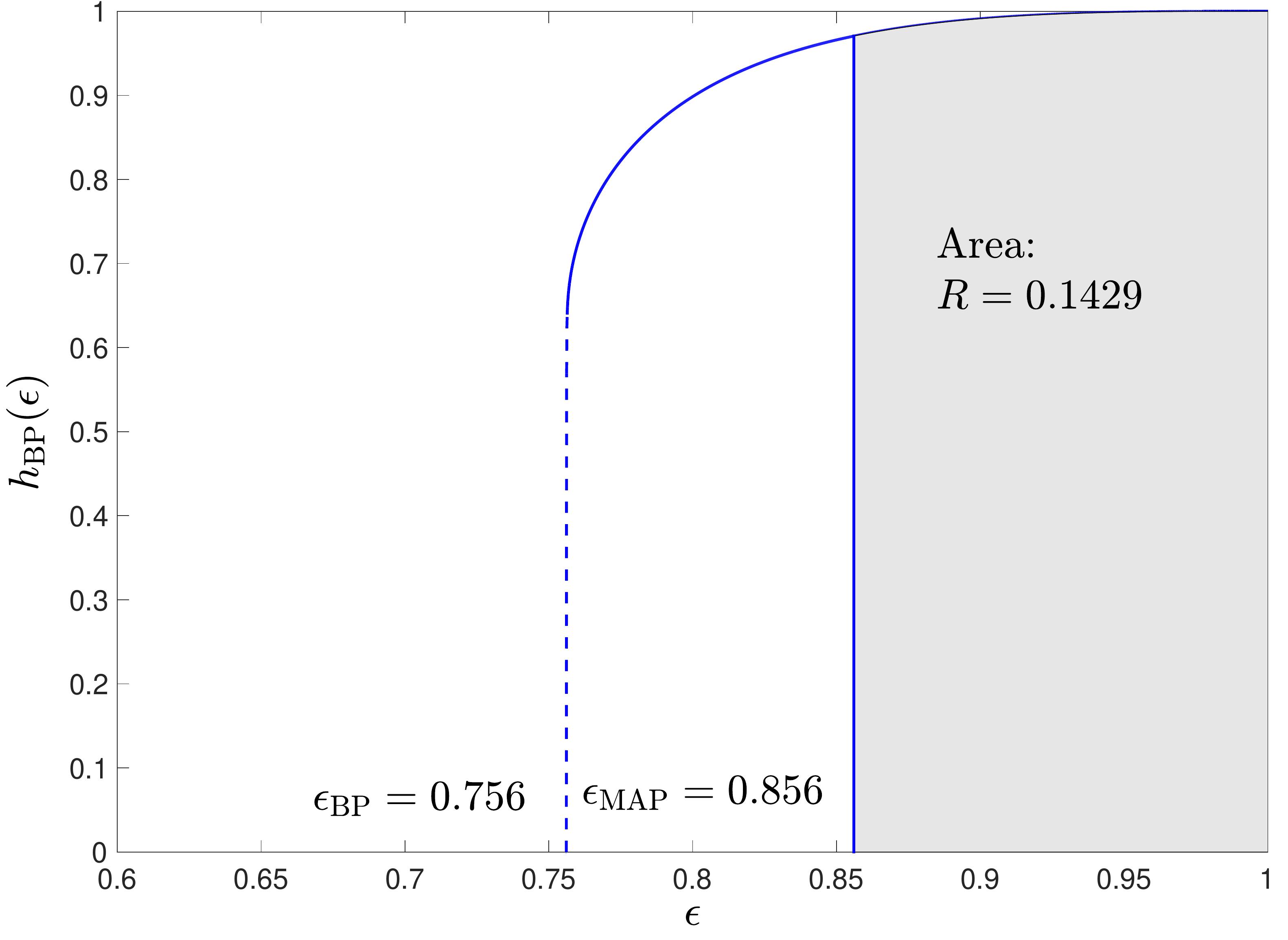}
\caption{BP EXIT function of a $(2,7)$-regular protograph-based GLDPC block code ensemble with $(7,4)$ Hamming component codes.} \label{fig:PG_EXITHam7}
\end{center}
\end{figure}

\begin{example} \label{ex:exit}Consider the $(2,7)$-regular protograph-based GLDPC block code ensemble with Hamming component codes of length $n^c=7$ from Example~\ref{ex:blockhamming}. The BP EXIT function $\hBP$ of this ensemble is shown in Fig.~\ref{fig:PG_EXITHam7}. The vertical line indicates the channel value at which the grey area below the curve is equal to the rate of the ensemble, which forms an upper bound  $\epsMAP=0.856$ on the threshold of an optimal MAP decoder. This follows from the area theorem \cite{akb04} and the fact that  $\hBP$ can never be below the EXIT function of the MAP decoder. In this case, the calculated BP threshold is given by $\epsBP=0.756$, and we see that there exists a large gap between the BP and the MAP thresholds, which indicates the suboptimality associated with BP decoding.\footnote{A detailed analysis of unstructured irregular  ensembles, including results on the tightness of this bound, can be found in \cite{mmu08}. For structured protograph ensembles, this technique has been applied in \cite{pvc+08}.}\hfill $\Box$\end{example}

\subsection{Threshold saturation of terminated SC-GLDPC code ensembles}
\begin{figure}
\begin{center}
\includegraphics[width=\columnwidth]{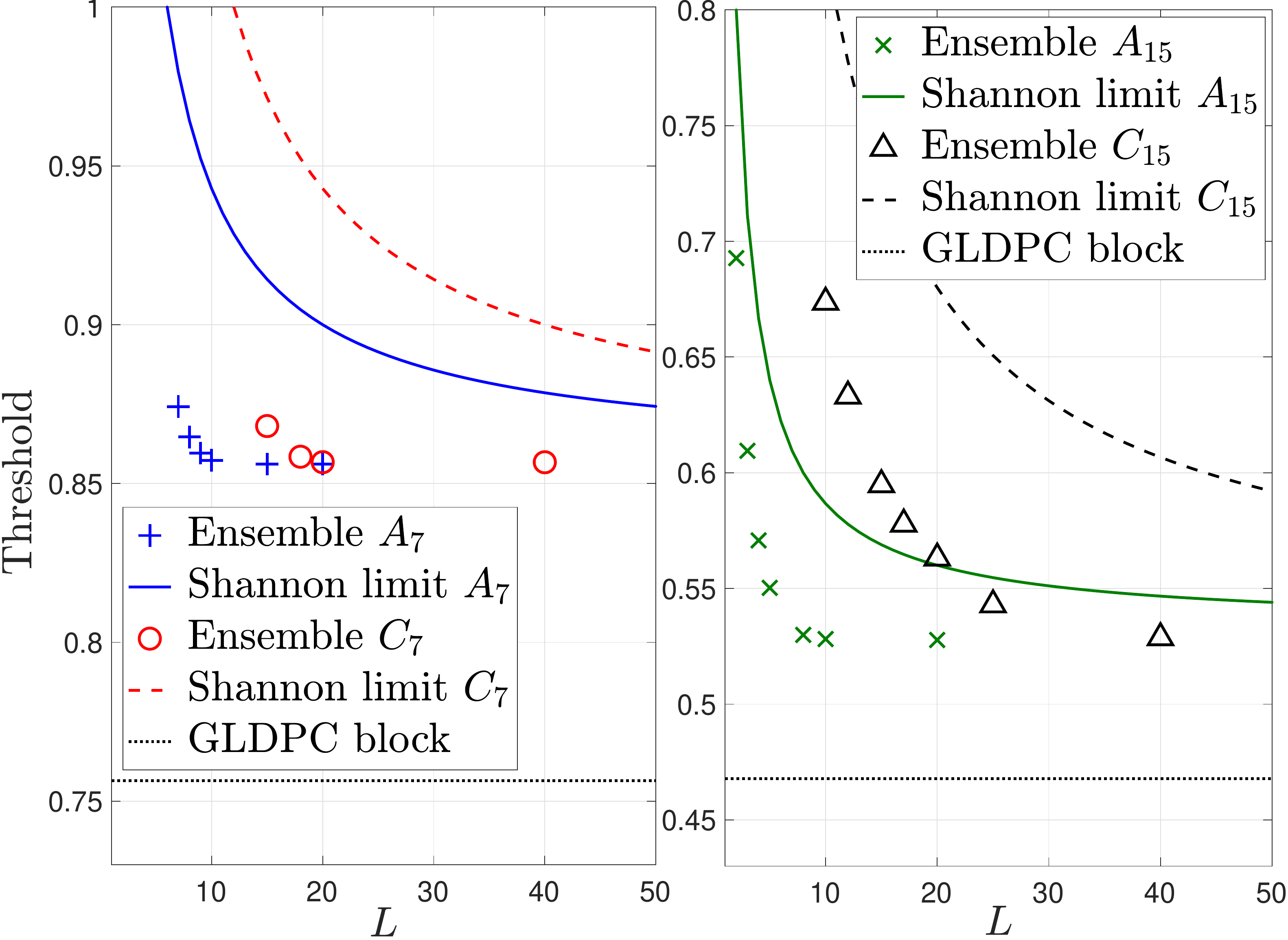}
\caption{BP decoding thresholds as functions of the coupling length $L$. Ensembles $A_7$ and $A_{15}$ are shown in comparison to ensembles $C_7$ and $C_{15}$, respectively. The dotted line indicates the thresholds of the underlying GLDPC block code ensembles.} \label{fig:ThresL}
\end{center}
\end{figure}

\begin{figure}
\begin{center}
\includegraphics[width=\columnwidth]{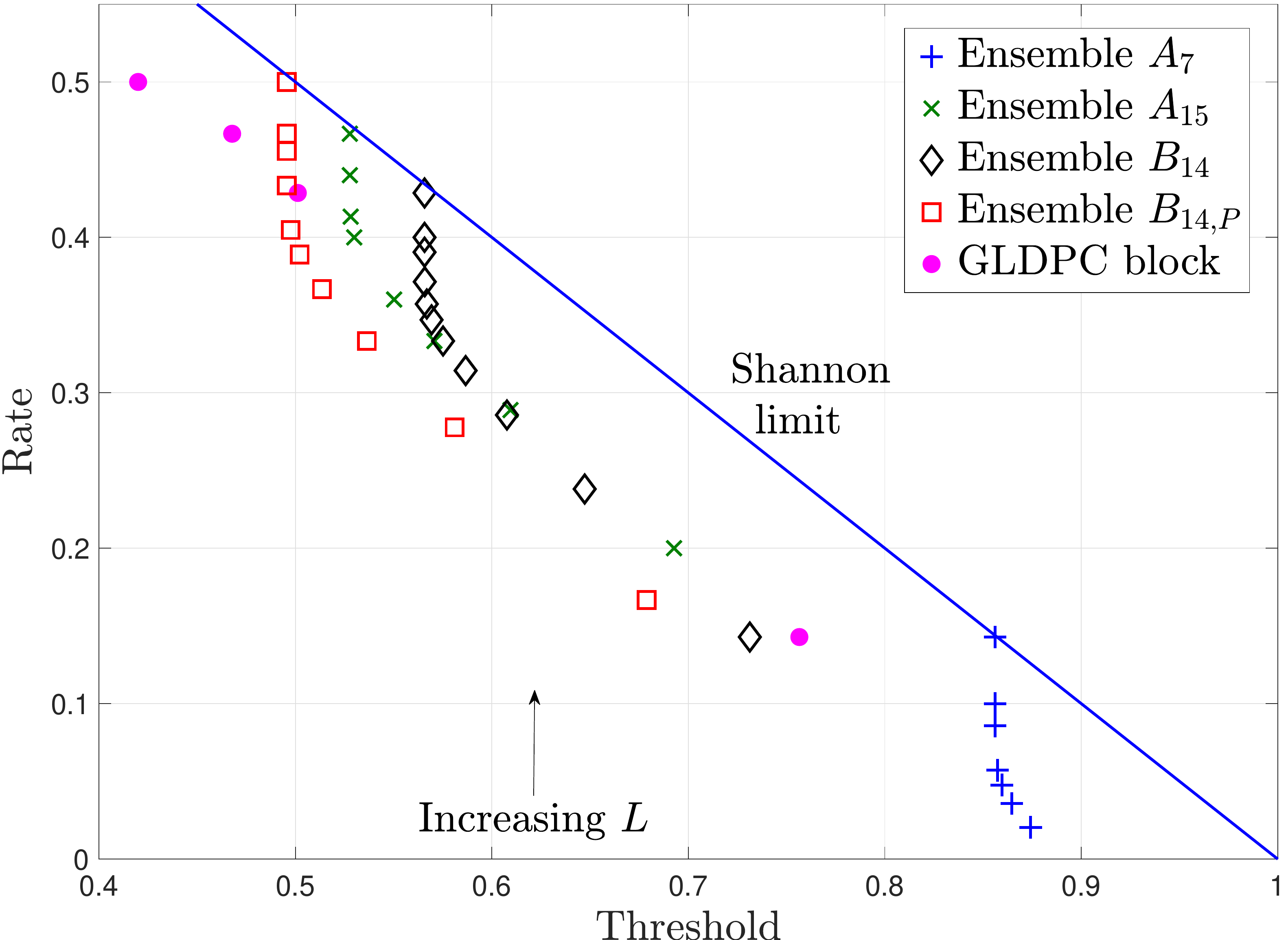}
\caption{BP decoding thresholds versus code rate for different $w=1$ SC-GLDPC code ensembles and coupling lengths $L$.} \label{fig:ThresRate}
\end{center}
\end{figure}

Assume now that we start encoding at time $t=0$ and terminate after $L$ time instants. As a result we obtain the terminated base matrix $\mathbf{B}_{[0,L-1]}$ from \eqref{termbase}. 
These terminated SC-GLDPC codes can be interpreted as GLDPC block codes that inherit the structure of convolutional codes. The length of these codes depends not only on the lifting factor $M$ but also on the coupling length $L$. For a fixed $L$, the BEC density evolution thresholds $\epsBP$  corresponding to codes with base matrix $\vec{B}_{[0,L-1]}$ can be calculated using the method described in Section \ref{sec:thressub}.  In Fig.~\ref{fig:ThresL}, the obtained thresholds for the $w=1$ ensembles $A_7$ and $A_{15}$ are compared with the BBC-based ensembles $C_7$ and $C_{15}$ for different coupling lengths $L$.  (The larger thresholds and Shannon limits of ensembles $C_7$ and $C_{15}$ compared to $A_7$ and $A_{15}$ for small to moderate $L$ is due to the fact that the larger $w$ BBC ensembles $C_7$ and $C_{15}$ have a larger rate increase parameter $\Delta$ than the $w=1$ ensembles $A_7$ and $A_{15}$. This difference vanishes as $L\rightarrow\infty$.) The thresholds of all the $w=1$ ensembles versus code rate are shown in Fig.~\ref{fig:ThresRate}.  Analogously to SC-LDPC codes (see \cite{mlc15}) with SPC constraints, it can be observed that, as $L \rightarrow \infty$, the BP thresholds numerically coincide with the upper bounds on the MAP decoding thresholds of the underlying block code ensembles, thus exhibiting the \emph{threshold saturation} phenomenon (see \cite{lscz10,kru11}).
 
%
%
%

%

\begin{figure}
\begin{center}
\includegraphics[width=\columnwidth]{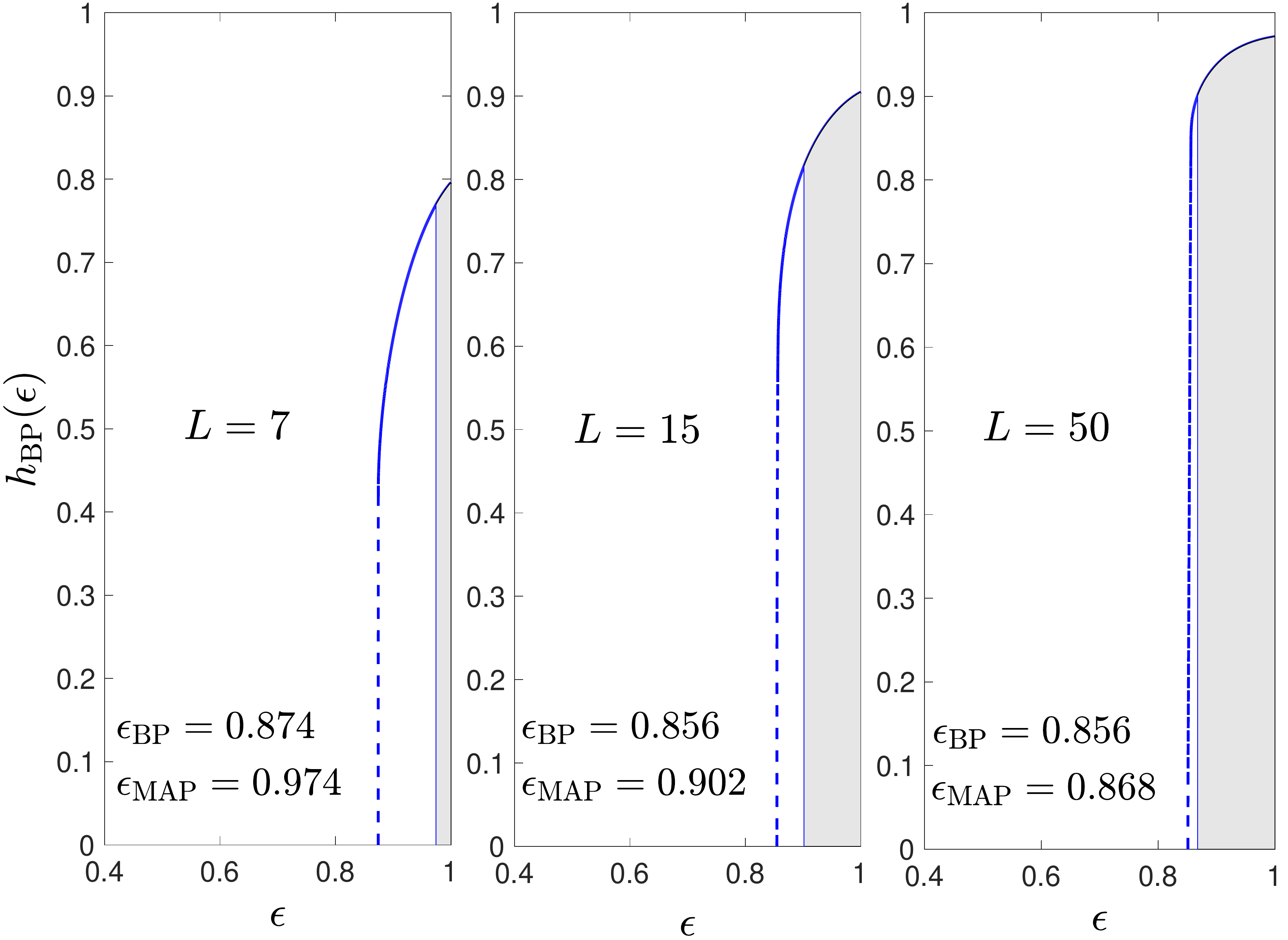}
\caption{BP EXIT functions of terminated SC-GLDPC codes from Ensemble $A_7$  for different coupling lengths $L$.} \label{fig:PG_EXITConv}
\end{center}
\end{figure}

The BP EXIT functions of the terminated codes from ensemble $A_7$ are shown in Fig.~\ref{fig:PG_EXITConv}, where it can be seen that, with increasing $L$, the BP and the MAP thresholds of the terminated SC-GLDPC code ensembles are converging. Moreover, the MAP thresholds (and hence the BP thresholds) of the terminated SC-GLDPC code ensemble can also be observed to converge to the MAP threshold $\epsMAP = 0.856$ of the underlying GLDPC block code ensemble with increasing $L${, demonstrating threshold saturation for the ensemble $A_7$. We note that the converging BP and MAP thresholds of the terminated SC-GLDPC code ensembles implies that, asymptotically, BP decoding of SC-GLDPC codes provides optimal (MAP) decoding performance.}

Further, we note that large values of $L$ are realistic in conjunction with sliding window decoders, like those suggested in \cite{ips+12}, where decoding delay and storage requirements depend on the window size $W$,  which is independent of the coupling length $L$ (typically $W \ll L$) of the transmitted code sequences. For shorter values of $L$, which induce rate loss, BP decoding of terminated SC-GLDPC codes is  suboptimal but still provides a flexible adjustment between code rate and threshold (see Fig.~\ref{fig:ThresRate}).  

\section{Distance Analysis of SC-GLDPC Codes}\label{sec:dist}
{In this section, we first perform an asymptotic minimum distance analysis of terminated and tail-biting SC-GLDPC code ensembles. We then present an approach to bound the free distance growth rate of the unterminated periodically time-varying SC-GLDPC code ensemble from above and below using the minimum distance growth rates of the terminated and tail-biting SC-GLDPC code ensembles, respectively, and provide numerical results. The ensemble free distance growth rate provides a measure of the strength of SC-GLDPC codes and an assessment of their ML decoding performance.}
\subsection{Minimum distance analysis of terminated SC-GLDPC codes}\label{sec:mindist}
From the convolutional protograph with base matrix $\mathbf{B}_{[0,\infty]}$ in \eqref{convbase}, we can form a periodically time-varying $M$-fold graph cover with period $T$ by applying the graph lifting operation described in Section~\ref{sec:proto} to the $b_c\times b_v$ submatrices $\mathbf{B}_0,\mathbf{B}_1,\ldots,\mathbf{B}_{w}$ in the first $T$ columns of $\mathbf{B}_{[0,\infty]}$ to form $Mb_c \times Mb_v$ submatrices $\mathbf{H}_0(t),\mathbf{H}_1(t+1),\ldots,\mathbf{H}_{w}(t+w)$, respectively, for {$t\in\{0,1,\ldots,T-1\}$}. These submatrices can then be repeated periodically (and indefinitely) to form a convolutional constraint matrix  $\mathbf{H}_{[0,\infty]}$ such that $\mathbf{H}_i(t+T)=\mathbf{H}_i(t)$, $\forall i,t$. An ensemble of periodically time-varying SC-GLDPC codes with period $T$, design rate $R=1-Mm^cb_c/Mb_v=1-m^cb_c/b_v$,\footnote{For simplicity, we again assume here the case that each generalized constraint node $c_i$ has $m^c$ independent parity checks.} and decoding constraint length $\nu_s=(w+1)Mb_v$ can then be derived by letting the permutation matrices used to form $\mathbf{H}_0(t),\mathbf{H}_1(t+1),\ldots,\mathbf{H}_{w}(t+w)$, for {$t\in\{0,1,\ldots,T-1\}$}, vary over all $M!$ choices of an $M\times M$ permutation matrix.

In \cite{adr11}, Abu-Surra, Divsalar, and Ryan presented a technique to calculate the average weight enumerator and asymptotic spectral shape function {$r(\delta)$} for protograph-based GLDPC block code ensembles. The spectral shape function can be used to test if an ensemble is \emph{asymptotically good}, \ie, if the minimum distance typical of most members of the ensemble is at least as large as $\delta_\mathrm{min}n$, where $\delta_\mathrm{min}$ is the \emph{minimum distance growth rate} of the ensemble and $n$ is the block length.\footnote{{Suppose that the first positive zero crossing of $r(\delta)$ occurs at $\delta = \delta_\mathrm{min}$. If $r(\delta)$ is negative in the range $0 < \delta < \delta_\mathrm{min}$, then $\delta_\mathrm{min}$ is called the minimum distance growth rate of the code ensemble, see }\cite{adr11}.}
\begin{example} Consider the $(2,7)$-regular GLDPC block code protograph with base matrix $\mathbf{B}$ from \eqref{hammingbase} and the generalized constraint nodes shown in Fig. \ref{fig:27ham}. If we suppose the constraint codes to be $(7,4)$ Hamming codes with parity-check matrix $\mathbf{H}_c$ from \eqref{hammingmatrices}, then the resulting  ensemble has design rate $R=1/7$, is asymptotically good, and has growth rate $\delta_\mathrm{min}=0.186$ \cite{adr11}. \hfill $\Box$\end{example}

We now consider the associated $(2,7)$-regular terminated SC-GLDPC code ensembles $A_7$ discussed above in Sections~\ref{sec:proto} and \ref{sec:thres}, whose protograph is shown in Fig.~\ref{fig:scprot}, with constraints corresponding to the $(7,4)$ Hamming code with parity-check matrix $\mathbf{H}_c$. After termination, the design rate of the ensemble is given by
\begin{equation}\label{ratesc}
R_L = 1-\frac{6(L+1)-2}{7L},
\end{equation}
where $\Delta=2$ in this case because the two leftmost (shortened) constraint nodes in Fig. \ref{fig:scprot} correspond to shortened codes with rate $1/3$, \ie, the number of parity checks in these two constraint nodes is $2$, while all the other constraint nodes have $m^c=3$ parity-checks. These ensembles were shown to have thresholds numerically indistinguishable from the MAP threshold of the underlying GLDPC block code ensemble as $L\rightarrow\infty$ in Section~\ref{sec:thres}.

{The asymptotic weight enumerator corresponding to a terminated or tail-biting convolutional protograph can be determined by applying the general approach of} \cite{adr11}{; however, we note that a useful conjecture regarding simplification of the numerical evaluation proposed in  }\cite{adr11} cannot be applied to SC ensembles. This conjecture relies on grouping together nodes of the same type and optimizing them {together, but} in the SC-GLDPC case, nodes from different time instants must be optimized separately, even if they are of the same type.

\begin{figure}[t]
\begin{center}
\includegraphics[width=\columnwidth]{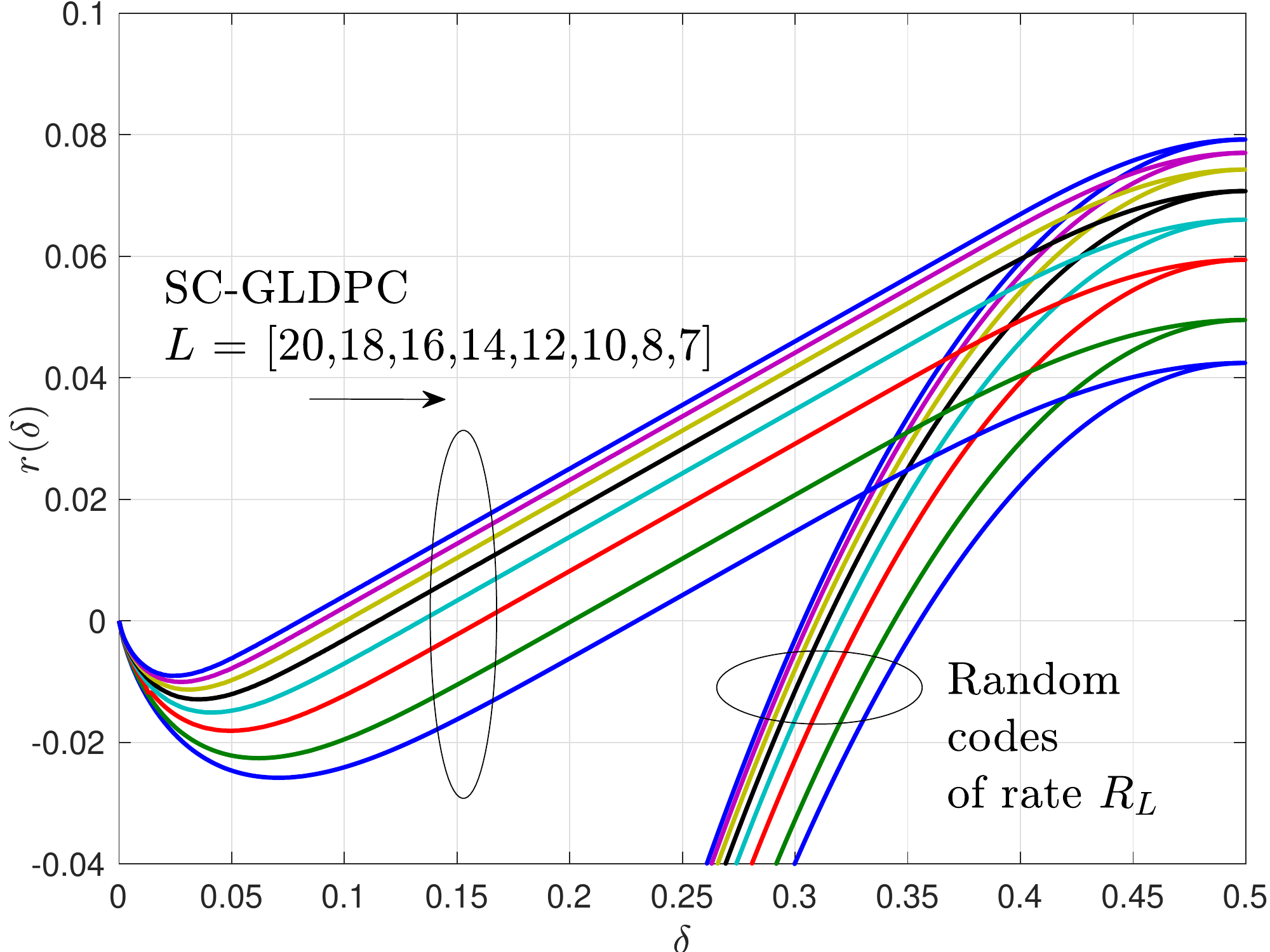}
\end{center}
\caption{Spectral shape functions of design rate $R_L$ SC-GLDPC code ensembles $A_7$ and random linear codes of the corresponding rate. {Curves from left to right in both groups correspond to $L=20,18,16,14,12,10,8,7$, respectively.}}\label{fig:termgrowth}
\end{figure}

Fig. \ref{fig:termgrowth} shows the asymptotic spectral shape functions for the SC-GLDPC code ensembles $A_7$ with coupling lengths {$L\in\{7,8,10,12,14,16,18,20\}$}. Also shown are the asymptotic spectral shape functions for {random} codes with the corresponding rates $R_L$ calculated using (see \cite{gal63})
\begin{equation}
r(\delta) = H(\delta)- (1-R_L)\ln(2),
\end{equation}
where $H(\delta)=-(1-\delta)\ln(1-\delta) - \delta\ln(\delta)$ is the binary entropy function. We observe that the SC-GLDPC code ensembles $A_7$ are asymptotically good and have relatively large minimum distance growth rates, ranging from about 25\% to 65\% of the (optimal) random coding growth rates. This indicates that long codes chosen from these ensembles have, with probability near one, a large minimum distance.  As $L$ increases, the design rate $R_L$ approaches $R=1/7$ and the minimum distance growth rate decreases, as was also observed in the case of SC-LDPC codes with SPC constraints (see \cite{mlc15}).

\subsection{Free distance analysis of SC-GLDPC code ensembles}\label{sec:freedist}
In Fig. \ref{fig:termgrowth} we saw that the minimum distance growth rates of terminated SC-GLDPC codes decrease as the coupling length $L$ increases. However, since SC-GLDPC codes can be decoded as unterminated (no preset coupling length) convolutional codes by employing a sliding window decoder, a more appropriate distance measure for assessing their ML decoding performance is the \emph{free distance} growth rate of the SC-GLDPC ensemble. In this section, we first calculate the minimum distance growth rates of TB-GLDPC code ensembles and show that, for sufficiently large coupling lengths, the growth rates coincide with those calculated for the terminated SC-GLDPC code ensembles in Section \ref{sec:mindist}. We then show that the growth rates of the TB-GLDPC code ensembles and the terminated SC-GLDPC code ensembles can be used to obtain lower and upper bounds on the free distance growth rate of the unterminated SC-GLDPC code ensemble, respectively.

\subsubsection{Minimum distance analysis of TB-GLDPC code ensembles}
\begin{figure}[t]
\begin{center}
\includegraphics[width=\columnwidth]{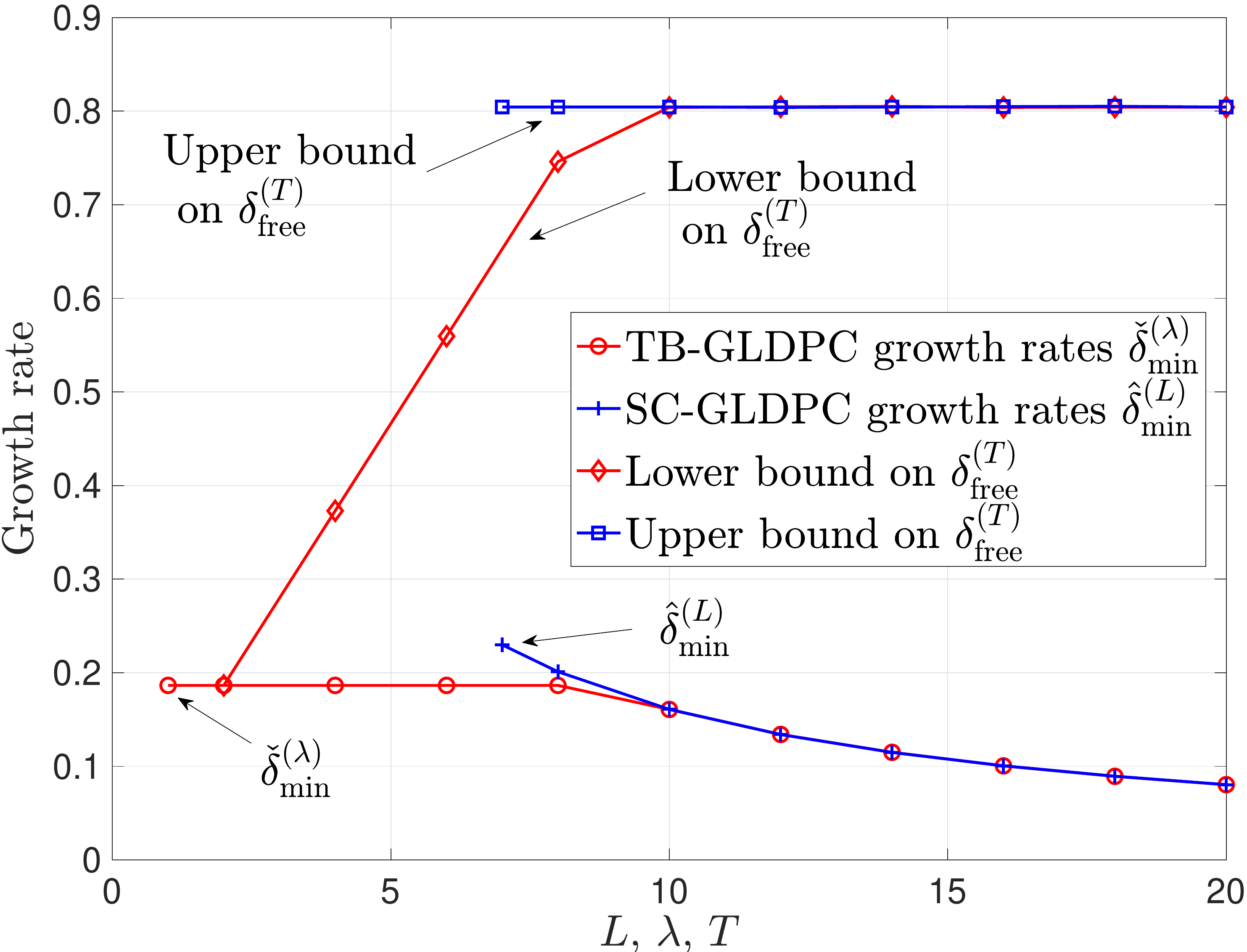}
\end{center}
\caption{Minimum distance growth rates of terminated SC-GLDPC code ensembles and TB-GLDPC code ensembles and calculated upper and lower bounds on the free distance growth rates of the associated periodically time-varying unterminated SC-GLDPC code ensembles.}\label{fig:termtbgrowth}
\end{figure}

We now consider terminating the protograph in Fig. \ref{fig:scprot} as a TB-GLDPC code with coupling length $\lambda$. Unlike the previous termination technique, this results in a $(2,7)$-regular protograph with design rate $R_\lambda = 1/7$ for all $\lambda$. The minimum distance growth rates {$\check{\delta}^{(\lambda)}_\mathrm{min}$} of the TB-GLDPC code ensembles are presented in Fig. \ref{fig:termtbgrowth} alongside those corresponding to the terminated SC-GLDPC code ensembles {$\hat{\delta}^{(L)}_\mathrm{min}$}. We observe that the TB-GLDPC growth rates remain constant at {$\check{\delta}^{(\lambda)}_\mathrm{min}=0.186$} (the growth rate of the original GLDPC block code ensemble {$\delta_\mathrm{min}$}) for $\lambda=1,2,\ldots,8,$ and then begin to decay to zero as $\lambda \rightarrow \infty$. Also, as a result of the convolutional structure, we observe that the TB-GLDPC and SC-GLDPC growth rates coincide for $L,\lambda\geq 10$. This is the same behavior observed for TB-LDPC and SC-LDPC codes with SPC constraints \cite{mpc13}. 

\subsubsection{Free distance bounds for SC-GLDPC code ensembles}
Now consider an ensemble of periodically time-varying unterminated SC-GLDPC codes with rate $R=1-b_cm^c/b_v$ and period $T$ constructed from a convolutional protograph with base matrix $\binf$ (see (\ref{convbase})) as described in Section \ref{sec:convproto}.  Using a modification of the proof techniques in \cite{mpc13,tzc10}, we can show that the average free distance of this ensemble is bounded below by the average minimum distance of an ensemble of TB-GLDPC codes derived from the base matrix $\btb$ (see (\ref{basetb})) with coupling length $\lambda=T$ . Here, we show that the average free distance of the unterminated SC-GLDPC ensemble can also be bounded above by the average minimum distance of the ensemble of terminated SC-GLDPC codes derived from the base matrix $\bterm$ (see (\ref{termbase})) with coupling length $L=T$.

\newtheorem{freedist}{Theorem}
\begin{freedist}
Consider a rate $R=1-b_cm^c/b_v$ unterminated, periodically
time-varying SC-GLDPC code ensemble with syndrome former memory $w$, decoding constraint length
$\nu_{s}=M(w+1)b_v$, and period $T$ derived from $\binf$. Let $\overline{d}^{(L)}_\mathrm{min}$ be
the average minimum distance of the terminated SC-GLDPC
code ensemble with block length $n=M b_vL$ and coupling length
$L$. Then the ensemble average free distance $\overline{d}_\mathrm{free}^{(T)}$ of the unterminated
SC-GLDPC code ensemble is bounded above by $\overline{d}_\mathrm{min}^{(L)}$ for termination
factor $L=T$, \ie,
\begin{equation}\label{dfreebound}
    \overline{d}_\mathrm{free}^{(T)} \leq \overline{d}^{(T)}_\mathrm{min}.
\end{equation}
\end{freedist}
\emph{Proof}. There is a one-to-one relationship between members of the periodically time-varying unterminated SC-GLDPC code ensemble and members of the corresponding terminated SC-GLDPC code ensemble with coupling length $L=T$. For any such pair of codes, every codeword $\mathbf{x}=[\begin{array}{cccc} x_0 & x_1 & \cdots & x_{M b_vL-1}\end{array}]$ in the terminated SC-GLDPC code can also be viewed as a codeword $\mathbf{x}_{[0,\infty]}=[\begin{array}{cccccc} x_0 & x_1 & \cdots & x_{M b_vL-1}&0&\cdots\end{array}]$ in the unterminated code. It follows that the free distance $d_\mathrm{free}^{(T)}$ of the unterminated code cannot be larger than the minimum distance $d_\mathrm{min}^{(T)}$ of the terminated code. The ensemble average result $\overline{d}_\mathrm{free}^{(T)} \leq \overline{d}^{(T)}_\mathrm{min}$ then follows directly. \hfill $\Box$

Since there is no danger of ambiguity, we will henceforth drop the overline notation when discussing ensemble average distance measures. 

\subsubsection{Free distance growth rates of SC-GLDPC code ensembles}\label{sec:growth}
For unterminated SC-GLDPC codes, it is natural to define the free distance growth rate with respect to the decoding constraint length $\nu_s$, \ie, as the ratio of the free distance $d_\mathrm{free}$ to $\nu_s$. By bounding $d_\mathrm{free}^{(T)}$ using (\ref{dfreebound}), we obtain an upper bound on the free distance growth rate as
\begin{equation}\label{ub}
\delta_\mathrm{free}^{(T)}=\frac{d_\mathrm{free}^{(T)}}{\nu_s} \leq \frac{\hat{\delta}_\mathrm{min}^{(T)}T}{(w+1)},
\end{equation}
where $\hat{\delta}_\mathrm{min}^{(T)} = {d_\mathrm{min}^{(T)}}/{n}={d_\mathrm{min}^{(T)}}/{(M  b_vT)}$ is the minimum distance growth rate of the terminated SC-GLDPC code ensemble with coupling length $L=T$ and base matrix $\mathbf{B}_{[0,T-1]}$. Further, using a similar argument to that presented in \cite{mpc13}, we have
\begin{equation}\label{lb}
\delta_\mathrm{free}^{(T)}\geq \frac{\check{\delta}_\mathrm{min}^{(T)}T}{(w+1)},
\end{equation}
where $\check{\delta}_\mathrm{min}^{(T)}$ is the minimum distance growth rate of the TB-GLDPC code ensemble with tail-biting coupling length $\lambda=T$ and base matrix $\btb$.

The free distance growth rate $\delta_\mathrm{free}^{(T)}$ that we bound from above using (\ref{ub}) is, by definition, an existence-type lower bound on the free distance of most members of the ensemble, \ie, with high probability a randomly chosen code from the ensemble has minimum free distance at least as large as ${\delta}_\mathrm{free}^{(T)}\nu_s$ as $\nu_s\rightarrow \infty$. 

\subsubsection{Numerical results}
As an example, we consider once more the $(2,7)$-regular SC-GLDPC code ensemble $A_7$ with memory $w=1$ and design rate $R=1/7$ depicted in Fig. \ref{fig:scprot}. For this case, we calculate the upper bound on the free distance growth rate of the periodically time-varying unterminated SC-GLDPC code ensemble as $\delta_\mathrm{free}^{(T)} \leq \hat{\delta}^{(T)}_\mathrm{min}T/2$ using (\ref{ub}) for coupling lengths $L=T\geq 7$. Fig. \ref{fig:termtbgrowth} displays the minimum distance growth rates $\hat{\delta}^{(L)}_\mathrm{min}$ of the terminated SC-GLDPC code ensembles defined by $\mathbf{B}_{[0,L-1]}$ for {$L\in\{7,8,10,12,\ldots,20\}$} that were calculated using the technique proposed in \cite{adr11} and the associated upper bounds on the unterminated SC-GLDPC code ensemble growth rates $\delta_\mathrm{free}^{(T)} \leq \hat{\delta}^{(T)}_\mathrm{min}T/2$ for $L=T$. Also shown are the minimum distance growth rates $\check{\delta}^{(\lambda)}_\mathrm{min}$ of the TB-GLDPC code ensembles defined by base matrix $\mathbf{B}^{(\lambda)}_{tb}$ for {$\lambda\in\{1,2,4,\ldots,20\}$} and the associated lower bounds on the unterminated SC-GLDPC code ensemble growth rates $\delta_\mathrm{free}^{(T)} \geq \check{\delta}^{(T)}_\mathrm{min}T/2$ for $\lambda=T$ calculated using (\ref{lb}).

We observe that the calculated TB-GLDPC code ensemble minimum distance growth rates $\check{\delta}_\mathrm{min}^{(\lambda)}$ remain constant for $\lambda=1, \ldots, 8$ and then start to decrease as the coupling length $\lambda$ grows, tending to zero as $\lambda$ tends to infinity. Correspondingly, as $\lambda$ exceeds $8$, the lower bound calculated for $\delta_\mathrm{free}^{(T)}$ levels off at $\delta_\mathrm{free}^{(T)} \geq 0.805$. The calculated terminated SC-GLDPC code ensemble minimum distance growth rates $\hat{\delta}_\mathrm{min}^{(L)}$ are larger for small values of $L$ (where the rate loss is larger) and decrease monotonically to zero as $L\rightarrow \infty$. Using (\ref{ub}) to obtain an upper bound on $\delta_\mathrm{free}^{(T)}$ we observe that, for $T\geq 10$, the upper and lower bounds coincide, indicating that, for these values of the period $T$, $\delta_\mathrm{free}^{(T)}=0.805$, significantly larger than the minimum distance growth rate $\delta_\mathrm{min}=0.186$ of the underlying GLDPC block code ensemble.\footnote{Note that the free distance growth rate may also be calculated with respect to the encoding constraint length $\nu_e$, which corresponds to the maximum number of transmitted symbols that can be affected by a single nonzero block of information digits. As a result of normalizing by the decoding constraint length, it is possible to have free distance growth rates larger than $0.5$. For further details, see \cite{mpc13}.} In addition, we note that, at the point where the upper and lower bounds on $\delta_\mathrm{free}^{(T)}$ coincide, the minimum distance growth rates for both termination methods also coincide. Recall that the TB-GLDPC code ensembles all have rate $1/7$, wheras the rate of the SC-GLDPC code ensembles is a function of the coupling length $L$ given by (\ref{ratesc}). {Although we have demonstrated the approach only for ensemble $A_7$, the general technique can be used to bound the free distance growth rate above and below for any regular or irregular periodically time-varying protograph-based unterminated SC-GLDPC code ensemble, thus allowing for the evaluation and comparison of different SC-GLDPC code designs from the perspective of minimum distance.}

While large free distance growth rates are indicative of good ML decoding performance, when predicting the iterative decoding performance of a code ensemble in the high SNR region other graphical objects such as trapping sets, pseudocodewords, absorbing sets, etc., come into effect. 
Based on results from the SPC case \cite{mpc13}, we would expect SC-GLDPC codes with large minimum/free distance growth rates to also have large trapping set growth rates, indicating good iterative decoding performance in the high SNR region.

\section{Finite-Length Analysis of GLDPC Block and SC-GLDPC Codes}\label{sec:finite}

To analyze the finite-length performance of LDPC codes over the BEC,  a decoding method called peeling decoding (PD) can be employed \cite{amru09}. PD is a simple algorithm that is initialized by removing all of the correctly received variable nodes, as well as their attached edges, from the Tanner graph of $\H$  after BEC transmission. The algorithm then iteratively removes degree-one check nodes from the graph, along with their attached variable nodes and edges. We now describe an extension of PD to GLDPC block and terminated SC-GLDPC codes, referred to as generalized peeling decoding (GPD). 

\subsection{Type vectors and initialization of GPD}
{Recall that}
\begin{align*}
E &= \{e^\mathrm{v}_{j,a}\,|\,j\in\{1,2,\ldots,n_v\}, a \in\{1,2,\ldots,\partial(v_j)\}\}\\
 &= \{e^\mathrm{c}_{i,b}\,|\,i\in\{1,2,\ldots,n_c\}, b \in\{1,2,\ldots,\partial(c_i)\}\}\\\
\end{align*}
{represents the set of edges in a given protograph}. In the Tanner graph of a resulting lifted constraint matrix $\H$ (resp. $\v{H}_{[0,L-1]}$), we say that a particular edge is of \emph{type} $e^\mathrm{v}_{j,a}$ if it connects a variable node and a constraint node that are copies of the two nodes that edge $e^\mathrm{v}_{j,a}$ connects in the protograph.  For a variable node $v$ in the Tanner graph of $\H$  (resp. $\v{H}_{[0,L-1]}$), we also define the \emph{variable node type} by a binary $|E|$-dimensional vector $\v{t}_v$, where each entry is a ``1" iff a particular edge type is connected to variable node $v$. Similarly, for a constraint node $c$, we define its type by a binary $|E|$-dimensional vector $\v{t}_c$. We denote the set of  variable and constraint node types in the Tanner graph of $\H$ (resp. $\v{H}_{[0,L-1]}$) by $\mathcal{F}_{\v{t}_v}$ and $\mathcal{F}_{\v{t}_c}$, respectively. Note that the sets $\mathcal{F}_{\v{t}_v}$ and $\mathcal{F}_{\v{t}_c}$ are determined from the connectivity of the protograph, as will be illustrated in Example~\ref{ex:edgetypes} below. Finally, we let $L_{\v{t}_v}$ and $R_{\v{t}_c}$ represent the number of variable and constraint nodes of  type $\v{t}_v$ and $\v{t}_c$ in $\v{H}$ (resp. $\v{H}_{[0,L-1]}$), respectively.

The details of the GPD algorithm are presented in Section \ref{GPD}.  We first discuss GPD initialization, which is identical to PD initialization: the correctly received variable nodes of $\H$ (resp. $\v{H}_{[0,L-1]}$) and their attached edges are removed from the graph. After initialization, the residual graph contains constraint nodes with types that are not included in $\mathcal{F}_{\v{t}_c}$, the set of constraint node types in the original graph, but the set of variable node types  $\mathcal{F}_{\v{t}_v}$ remains the same. We now define $\mathcal{D}(\v{t}_c)$ as the set of constraint node types that can appear in the graph of $\H$ (resp. $\v{H}_{[0,L-1]}$) after GPD initalization when a constraint node of type $\v{t}_c\in \mathcal{F}_{\v{t}_c}$ loses one or more edges. The extended set of all possible constraint node types which are present in the residual graph after  GPD initialization is then given by $\overline{\mathcal{F}}_{\v{t}_c}=\bigcup_{\v{t}_c\in\mathcal{F}_{\v{t}_c}}\mathcal{D}(\v{t}_c)$. 

\begin{example} \label{ex:edgetypes}
To illustrate the type vectors, consider the $(2,7)$-regular GLDPC block code protograph of Example \ref{ex:blockhamming}. In this case, if we order the protograph edges as $$(e^\mathrm{c}_{1,1},e^\mathrm{c}_{1,2},\ldots,e^\mathrm{c}_{1,7},e^\mathrm{c}_{2,1},e^\mathrm{c}_{2,2},\ldots,e^\mathrm{c}_{2,7}),$$ then the type vector for $c_1$ is $\v{t}_{c_1} = (1,1,\ldots,1,0,0,\ldots,0) \in \mathcal{F}_{\v{t}_c}$ and the type vector for $c_2$ is $\v{t}_{c_2} = (0,0,\ldots,0,1,1,\ldots,1)\in \mathcal{F}_{\v{t}_c}$.\footnote{The ordering of the protograph edges (which can affect the entries of the type vector) does not matter provided that we use a consistent ordering for all node types.} Both vectors are of length $|E| = 14$, the number of edges in the protograph, and have weight $n^c=7$, the length of the constraint code. In any lifted graph with lifting factor $M$, there are precisely $M$ copies of each edge, variable node, or constraint node of a given type, where the types are defined from the protograph as described above. Corresponding to each of the constraint node types in this example, $2^7=128$ residual types can appear in the graph when edges are removed. Thus, in total, $\overline{\mathcal{F}}_{\v{t}_c}$ contains  256 constraint node types. \hfill $\Box$\end{example}

%
%
%
%
%

%
%
%

According to the above definitions, the expected degree distribution (DD) of the residual graph after   initialization can be expressed as follows:
\begin{align}\label{eq:L0}
L_{\v{t}_v}(0)&=\epsilon L_{\v{t}_v}, \\\label{eq:R0}
R_{\v{t}^\prime_c}(0)&=\sum_{\substack{\v{t}_c\in\mathcal{F}_{\v{t}_c}\\ \v{t}_c^\prime\in \mathcal{D}(\v{t}_c)}}R_{\v{t}_c} \epsilon^{|\v{t}_c^\prime|}(1-\epsilon)^{|\v{t}_c|-|\v{t}_c^\prime|},
\end{align}
for $\v{t}_v\in\mathcal{F}_{\v{t}_v}$ and $\v{t}^\prime_c\in\overline{\mathcal{F}}_{\v{t}_c}$, where  $L_{\v{t}_v}(0)$ (resp. $R_{\v{t}^\prime_c}(0)$) represents the number of variable (resp. constraint) nodes of type $\v{t}_v$ (resp. $\v{t}^\prime_c$) after GPD initialization and $|\v{t}^\prime_c|$ (resp. $|\v{t}_c|$) is the weight of the vector $\v{t}^\prime_c$ (resp. $\v{t}_c$).

\subsection{Decodable constraint nodes and the GPD}\label{GPD}


In general, each constraint node type in the protograph $\B$ (resp. $\v{B}_{[0,L-1]}$) of a GLDPC code can be associated with a different constraint code.  Let $\mathcal{C}_{\v{t}_c}$ be the constraint code associated with the constraint nodes in the base matrix $\B$ (resp. $\v{B}_{[0,L-1]}$) of type $\v{t}_c\in\mathcal{F}_{\v{t}_c}$. By extension, each constraint node in the graph of $\H$ (resp. $\v{H}_{[0,L-1]}$) is associated with a constraint code according to its type. 
After GPD initialization, the type of a given constraint node can be modified from $\v{t}_c$ to $\v{t}_c^\prime$, where $|\v{t}_c^\prime|<|\v{t}_c|$, and we say that $\v{t}_c^\prime$ is the input erasure pattern seen by the constraint code $\mathcal{C}_{\v{t}_c}$. 
The question now is if, by decoding the constraint code $\mathcal{C}_{\v{t}_c}$ associated with a constraint node of type $\v{t}_c^\prime$ in the residual graph using a given decoding algorithm, for instance ML decoding or some suboptimal algorithm, we are able to recover the $|\v{t}_c^\prime|$ variables still connected to the constraint node. In general, for each constraint code  $\mathcal{C}_{\v{t}_c}$, only a subset of input erasure patterns can be decoded. This subset is denoted by $\mathcal{A}(\v{t}_c)\subset\mathcal{D}(\v{t}_c)$.  If  a constraint node in the residual graph is of type $\v{t}_c^\prime\in\mathcal{A}(\v{t}_c)$, then we say it is a \emph{decodable} constraint node and  $\v{t}_c^\prime$ a \emph{decodable} constraint node type. For example, if all the constraint codes are SPCs, then only constraint node types with input erasure patterns containing exactly one erasure are decodable. However, if the constraint code is a $(7,4)$ Hamming code with ML decoding, then all input erasure patterns with one and two erasures and some input erasure patterns of weight three are decodable.

The set of all decodable constraint node types is defined as $\mathcal{A}\doteq\cup_{\v{t}_c\in\mathcal{F}_{\v{t}_c}}\mathcal{A}(\v{t}_c)\subset\overline{\mathcal{F}}_{\v{t}_c}$. Given the discussion above, the GPD algorithm can now be seen as a  straightforward extension of PD for LDPC codes to GLDPC codes. After the graph is initialized, GPD chooses one constraint node at random from the graph that is decodable. This constraint node, all connected variable nodes, and all attached edges are then removed from the graph.  GPD continues in this way until there are no further constraint nodes that can be removed from the graph, which corresponds to a \emph{decoding failure}, or until there are no variable nodes left in the graph, which corresponds to \emph{successful decoding}.

\subsection{Expected graph evolution}\label{expectedgraph}
We now define the normalized DD at time $\tau$ as 
\begin{align}\label{eq:norm}
\tau\doteq\frac{\ell}{n}, \qquad r_{\v{t}_c}(\tau)\doteq\frac{R_{\v{t}_c}(\tau)}{n}, \qquad \l_{\v{t}_v}(\tau)\doteq\frac{\L_{\v{t}_v}(\tau)}{n},
\end{align} 
where $\ell$ is the GPD iteration index, $R_{\v{t}_c}(\tau)$ (resp. $L_{\v{t}_v}(\tau)$) is the number of constraint (resp. variable) nodes in the graph of type $\v{t}_c$ (resp. $\v{t}_v$) at time $\tau$,  and $n=Mb_v$ (resp. $Mb_vL$) is the block (resp. termination) length. 

{In }\cite{lmss01b}{, it is shown that if we apply PD to elements of an LDPC code ensemble, then the expected DD of the sequence of residual graphs can be described as the solution of a set of differential equations. This analysis is based on a result on the evolution of Markov processes due to Wormald} \cite{worm95}{. Furthermore, the deviation of the process w.r.t. the expected evolution decreases exponentially fast with the LDPC code block length, and this result was used in} \cite{amru09} {to analyze the finite-length BEC performance of LDPC block codes. This methodology has been extended to unstructured GLDPC codes in} \cite{lok19} {and also to spatially coupled codes with split-component codes} \cite{ztk18}{, which can be consider as a particular sub-class of SC-GLDPC codes. In a similar way, we can investigate the  finite-length BEC performance of  GLDPC codes constructed from protographs by analyzing the statistical evolution of the normalized DD  in} \eqref{eq:norm} {during the decoding process. As shown in} \cite{amru09}{, the expected value of $r_{\v{t}_c}(\tau)$ and $\l_{\v{t}_v}(\tau)$, denoted by  $\hat{r}_{\v{t}_c}(\tau)$ and $\hat{\l}_{\v{t}_v}(\tau)$, respectively, can be computed as the solution to the following system of differential equations}
\begin{align}
&\frac{\partial \hat{\l}_{\v{t}_v}(\tau)}{\partial \tau}=\nonumber\\&\E[L_{\v{t}_v}(\tau+\frac{1}{n})-L_{\v{t}_v}(\tau)\Big|\{\hat{l}_{\v{t}_v}(\tau),\hat{r}_{\v{t}_c}(\tau)\}_{\v{t}_v\in\mathcal{F}_{\v{t}_v}, \v{t}_c\in\overline{\mathcal{F}}_{\v{t}_c}}] \label{eq:system1},
\\
&\frac{\partial \hat{r}_{\v{t}_c}(\tau)}{\partial \tau}=\nonumber\\&\E[R_{\v{t}_c}(\tau+\frac{1}{n})-R_{\v{t}_c}(\tau)\Big|\{\hat{l}_{\v{t}_v}(\tau),\hat{r}_{\v{t}_c}(\tau)\}_{\v{t}_v\in\mathcal{F}_{\v{t}_v}, \v{t}_c\in\overline{\mathcal{F}}_{\v{t}_c}}] \label{eq:system2},
\end{align}
\ie,  the derivative of $\hat{r}_{\v{t}_c}(\tau)$ w.r.t. $\tau$ in \eqref{eq:system2} can be evaluated by computing the variation in the number of constraint nodes of type $\v{t}_c$ with GPD iteration given that the normalized DD at time $\tau$ is  at its mean $\{\hat{\l}_{\v{t}_v}(\tau),\hat{r}_{\v{t}_c}(\tau)\}_{\v{t}_v\in \mathcal{F}_{\v{t}_v}, \v{t}_c\in \overline{\mathcal{F}}_{\v{t}_c}}$.\footnote{{This result, based on Wormald's theorem} \cite{worm95}{, requires certain conditions on the random sequence of GLDPC Tanner graphs during GPD to be met. We do not include the proof that these conditions are met, but an equivalent proof for unstructured GLDPC codes can be found in} \cite{lok19}.}
 A similar interpretation holds for \eqref{eq:system1}.  Further, the solution to \eqref{eq:system1} and \eqref{eq:system2} is unique and, with probability \mbox{$1-\mathcal{O}(\text{e}^{-\sqrt{n}})$}, any particular realization of the normalized DD in \eqref{eq:norm} deviates from its mean by a factor of less than $n^{-1/6}$ for the initial conditions
\begin{align}\label{eq:initmean1}
\hat{r}_{\v{t}_c}(0)&=\E[r_{\v{t}_c}(\ell=0)]=\E[R_{\v{t}_c}(\ell=0)]/n, \\\label{eq:initmean2}
\hat{\l}_{\v{t}_v}(0)&=\E[\l_{\v{t}_v}(\ell=0)]=\E[\L_{\v{t}_v}(\ell=0)]/n,
\end{align} 
which can be computed from \eqref{eq:L0} and \eqref{eq:R0} \cite{amru09}. The actual computation of the expectations in \eqref{eq:system1} and \eqref{eq:system2} is described in \cite{omc15}. 
%
%
%
%
%
The GPD threshold is defined as the maximum  value of $\epsilon$ for which the expected  fraction of decodable constraint nodes 
\begin{align}\label{eq:r1mean}
\hat{a}(\tau)\doteq\sum_{\v{t}_c\in\mathcal{A}} \hat{r}_{\v{t}_c}(\tau)
\end{align}
is positive for any $\tau\in[0, \epsilon)$, where $\hat{a}(\tau)$ is the mean of the random process
\begin{align}\label{eq:c1}
a(\tau)\doteq\sum_{\v{t}_c\in\mathcal{A}} r_{\v{t}_c}(\tau).
\end{align}
Finally, we can compute the expected fraction of variable nodes in the graph at any time $\tau$, denoted by $\hat{v}(\tau)$, as 
\begin{align}\label{eq:c1-2}
\hat{v}(\tau)\doteq\sum_{\v{t}_v \in \mathcal{F}_{\v{t}_v}} \hat{l}_{\v{t}_v}(\tau).
\end{align}

In addition to characterizing the asymptotic behavior, \ie, to computing the GPD threshold ensemble, the solution to the system of equations given by \eqref{eq:system1} and \eqref{eq:system2} can be used to determine the quantities needed to assess the finite-length performance of GLDPC block and terminated SC-GLDPC codes. We refer to \emph{critical points} as the points in time for which $\hat{a}(\tau)$ has a local minima.
As shown in \cite{amru09}, the average (over the ensemble of codes) error probability is dominated by the probability that the process $a(\tau)$ \emph{survives}, \ie, does not go to zero  around the critical points.
Therefore, characterizing the critical points and the expected fraction of decodable constraint nodes in the graph at those points in time are the parameters needed to determine the GLDPC block or terminated SC-GLDPC code finite-length performance, and they can be computed from \eqref{eq:system1} and \eqref{eq:system2}. 
%

\subsection{Numerical results: GLDPC  block codes}\label{sec:finiteSC}

With the tools described above, we can now investigate the asymptotic and finite-length performance of GLDPC code ensembles.  We start by considering the uncoupled $(2,7)$-regular GLDPC block code ensemble from Example~\ref{ex:blockhamming}.

\begin{figure}[t]
\centering \includegraphics[width=\columnwidth]{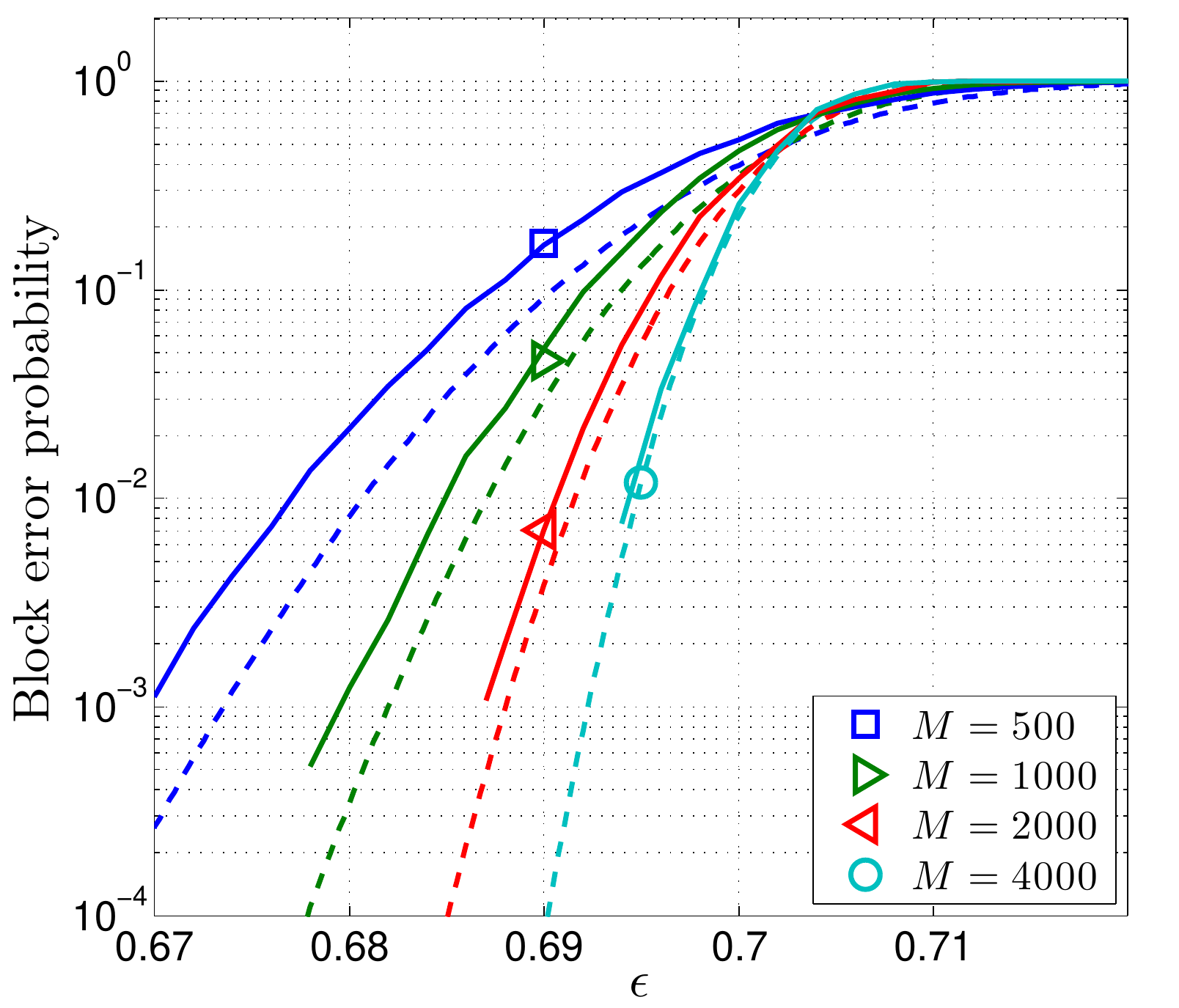}
\caption{Simulated performance (solid lines) and   estimated performance using \eqref{SL2} (dashed lines) for the $(2,7)$-regular GLDPC block code ensemble with $(7,4)$ Hamming  constraint codes and an ML decoder.}\label{27ML_sims} 
\end{figure}

\begin{example} \label{ex:blockscaling}Consider the $(2,7)$-regular GLDPC block code ensemble of Example 1. Assume  that a $(7,4)$ Hamming  code is associated with each  of the two constraint nodes and that the constraint codes are decoded using ML decoding. The design rate of this ensemble is $R=1/7$. All constraint node types with one or two erasures can be decoded, as well as  some constraint node types with three erasures.
Fig. \ref{27ML} shows the evolution of the expected fraction of decodable constraint nodes $\hat{a}(\tau)$ versus the expected fraction of variable nodes $\hat{v}(\tau)$  in the graph for different $\epsilon$ values.\footnote{Note that the time variable $\tau$ runs backwards in this figure (right to left), in the sense that small values of $\tau$ correspond to $\hat{v}(\tau)$ on the right, where the graph still contains a relatively large fraction of variable nodes, whereas large values of $\tau$ correspond to small values of  $\hat{v}(\tau)$ on the left.} We also include a set of 10 simulated trajectories of $a(\tau)$ for $\epsilon=0.69$ to demonstrate that they concentrate around the predicted mean. Note first that $\hat{a}(\tau)$ has a single critical point at $\hat{v}(\tau^*)\approx 0.43$. Indeed, we can compute the threshold $\epsilon^*$ as the maximum value of $\epsilon$  for which the minimum is exactly zero, and in this case we obtain $\epsilon^*\approx 0.7025$.

\begin{figure}[t]
\centering \includegraphics[width=\columnwidth]{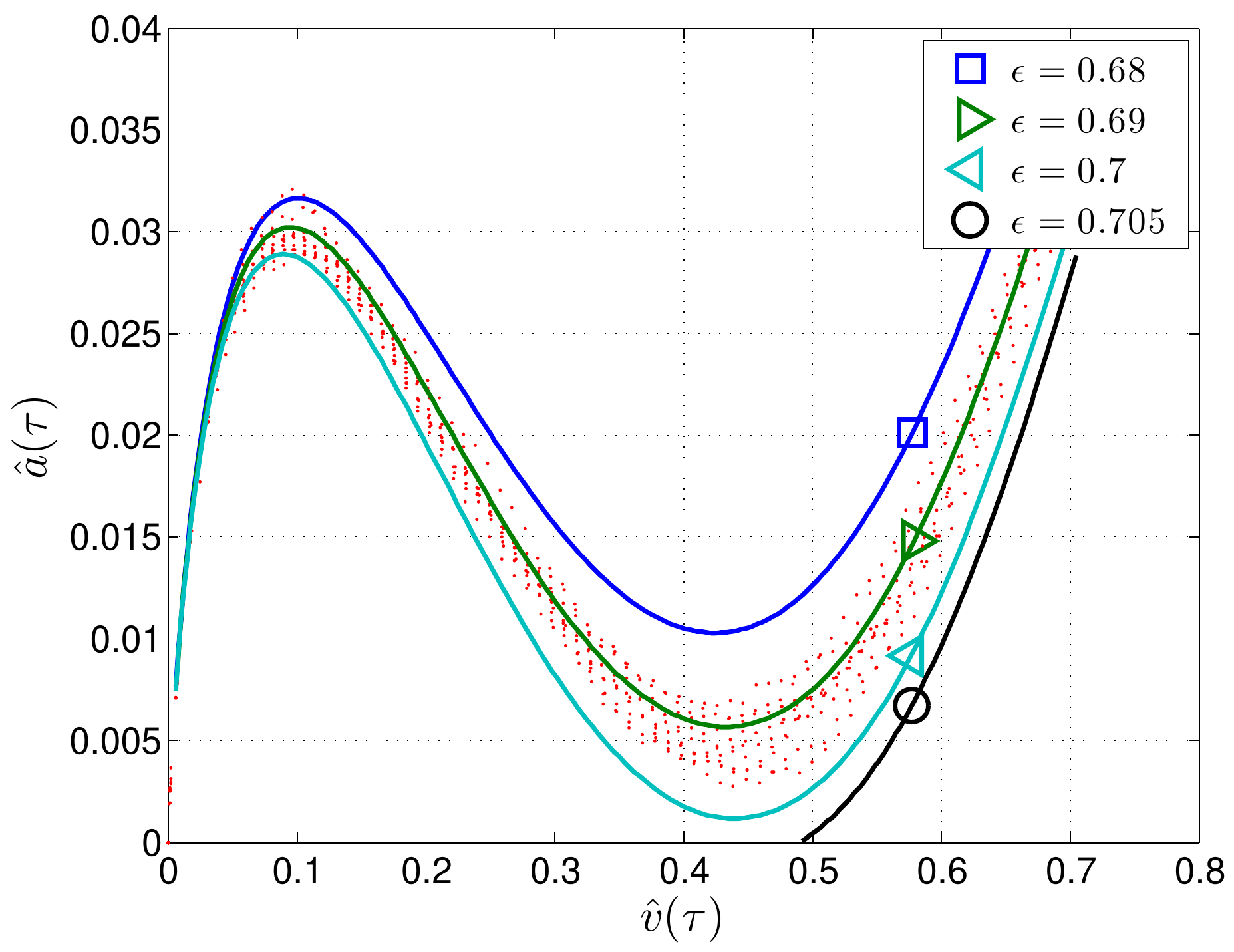}
\caption{Evolution of the expected fraction of decodable constraint nodes $\hat{a}(\tau)$ in the residual graph during iterations of the GPD for the $(2,7)$-regular GLDPC block code ensemble with $(7,4)$ Hamming constraint codes and an ML decoder. The dotted curves represent simulated trajectories computed for  $\epsilon=0.69$ with lifting factor $M=4000$.}\label{27ML} 
\end{figure}

The finite-length error probability is dominated by the statistics of $a(\tau)$ around $\tau^*$.  Following \cite{amru09}, for each $n$ and $\epsilon$ pair, we can estimate the finite-length error probability {as}
\begin{align}\label{SL}
P_{\text{Block}}=\text{Pr}(a(\tau)=0, v(\tau)>0)\approx \mathcal{Q}\left(\frac{\hat{a}(\tau^*)}{\sqrt{\text{Var}(a[\tau^*])}}\right),
\end{align}
{where}  $\hat{a}(\tau^*)$ is the expected value of $a(\tau)$ at $\tau^*$ and $\text{Var}[a(\tau^*)]$ represents its variance. $\hat{a}(\tau^*)$ was computed using numerical integration of the system of differential equations in \eqref{eq:system1}-\eqref{eq:system2}, and $\text{Var}(a[\tau^*])$ was estimated using Monte Carlo simulation.\footnote{$\text{Var}(a[\tau^*])$  can be obtained from the solution of the \emph{covariance evolution} system of differential equations, first presented  in \cite{amru09} for  LDPC code ensembles}  Also in \cite{amru09}, the authors showed  that the ratio of the expected number of degree-one constraint nodes to the standard deviation at the critical point approximately scales as $\alpha\sqrt{n}(\epsilon^*-\epsilon)$, where $\alpha$ is a \emph{scaling parameter} that only depends on the DD. In the GLDPC case, simulated trajectories for $a(\tau)$ suggest that the same scaling holds and that the  performance  for any pair $(n,\epsilon)$ can be estimated {as} 
\begin{align}\label{SL2}
P_{\text{Block}}\approx  \mathcal{Q}\left(\alpha\sqrt{n}(\epsilon^*-\epsilon)\right).
\end{align}

{After} computing $\hat{a}(\tau^*)/\sqrt{\text{Var}(a(\tau^*))}$ for  a given $(n,\epsilon)$ pair, we estimate $\alpha$  by equating the arguments in \eqref{SL} and \eqref{SL2}, so that
\begin{align}
\alpha = \frac{1}{\sqrt{n}(\epsilon^*-\epsilon)}\frac{\hat{a}(\tau^*)}{\sqrt{\text{Var}(a[\tau^*])}}.
\end{align}
For  the $(2,7)$-regular GLDPC block  code ensemble with $(7,4)$ Hamming  constraint codes and an ML decoder, we thus obtain $\alpha=1.8024$. In Fig.~\ref{27ML_sims}, we plot the simulated performance versus \eqref{SL2}, where we observe that the estimate is very accurate for a sufficiently large lifting factor. \hfill$\Box$\end{example}

\begin{figure}[h!]
\centering \includegraphics[width = \columnwidth]{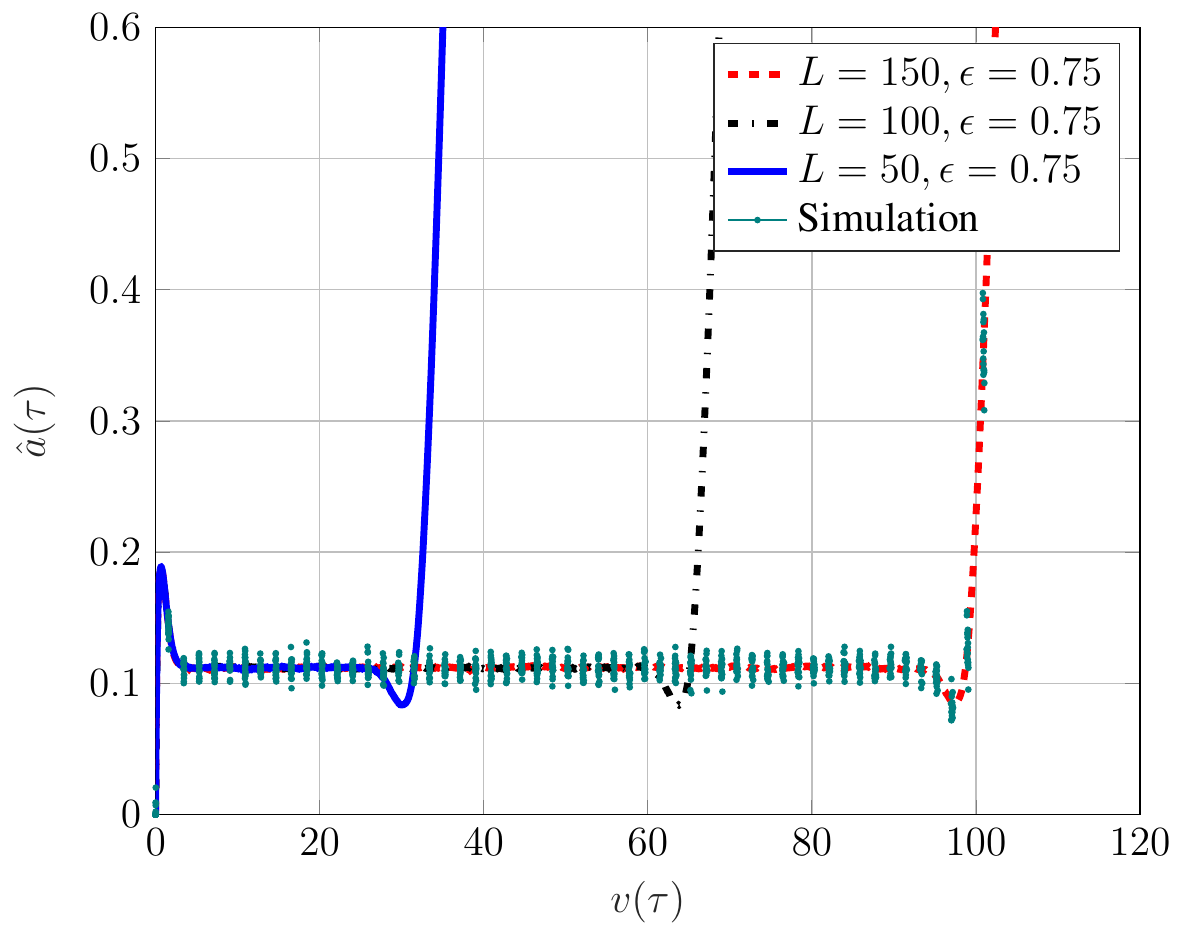}
\caption{Evolution of the expected fraction of decodable constraint nodes $\hat{a}(\tau)$ in the residual graph during iterations of the GPD for the terminated $(2,7)$-regular $A_7$ SC-GLDPC code ensemble with coupling lengths $L=50$, $100$, and $150$, $(7,4)$ Hamming constraint codes, and  an ML decoder. 
The dotted curves represent simulated {trajectories for}  $L=150$ and lifting factor $M=1000$. 
}
\label{27ML_simsv2} 
\end{figure}


\subsection{Numerical results: SC-GLDPC  codes}\label{sec:finiteSC}
We now investigate the asymptotic and finite-length performance of a terminated version of the coupled $A_7$ SC-GLDPC code ensemble from Example~\ref{ex:convhamming}.

\begin{figure}[t!]
\centering \includegraphics[width = \columnwidth]{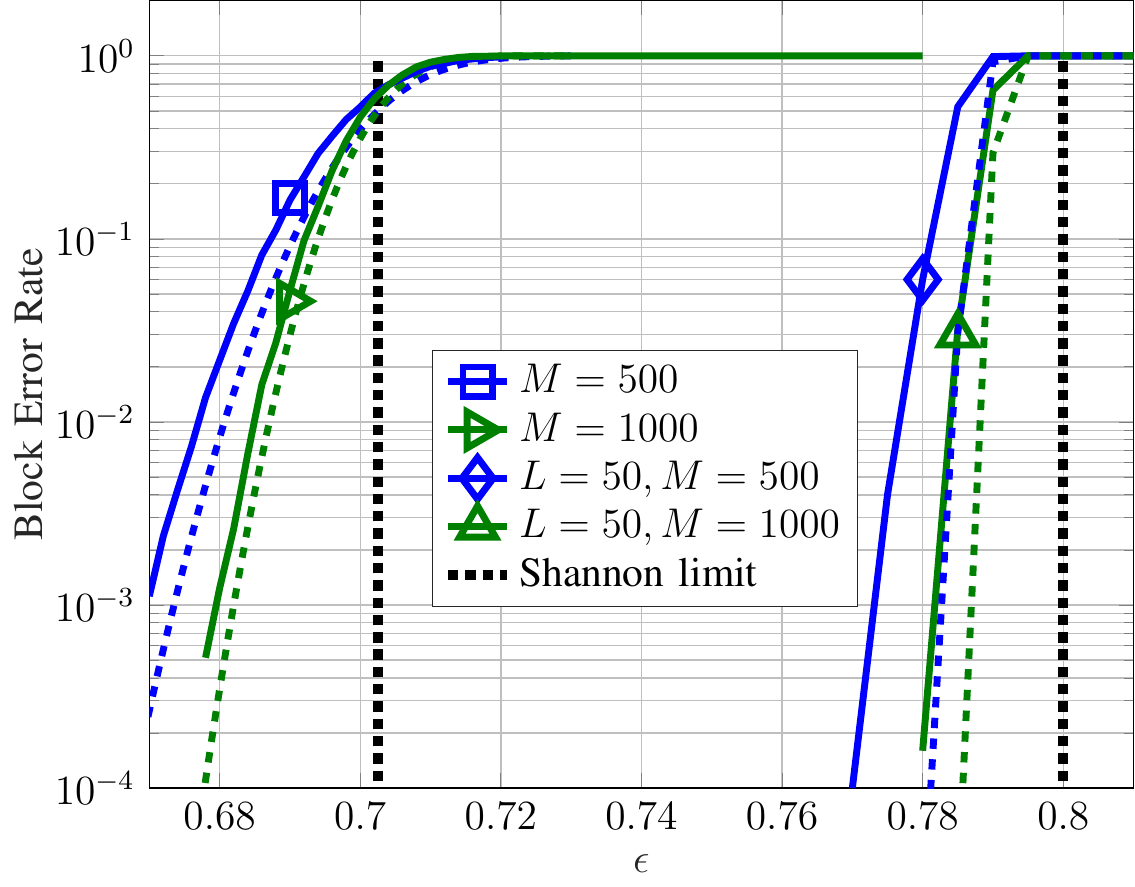}
\caption{Simulated performance (solid lines) and  estimated performance (dashed lines) for  the $(2,7)$-regular GLDPC block code and the terminated $(2,7)$-regular $A_7$ SC-GLDPC code ensembles with $M=500$ and $M=1000$.}
\label{27ML_simsv3} 
\end{figure}

\begin{example}Following a similar procedure as in Example~\ref{ex:blockscaling}, we now analyze the finite-length behavior of terminated SC-GLDPC codes.  In Fig.~\ref{27ML_simsv2}, we show the evolution of the expected fraction $\hat{a}(\tau)$ of decodable check nodes during iterations of the GPD for a terminated version of the $A_7$ SC-GLDPC code ensemble of Example~\ref{ex:convhamming} (corresponding to the GLDPC block code ensemble of Example~\ref{ex:blockhamming}) with  coupling lengths $L=50$, $100$, and $150$, lifting factor $M=1000$, and a channel parameter $\epsilon=0.75$. {Also included is a set of simulated decoding trajectories of $a(\tau)$, computed for $L=150$}. Unlike the GLDPC block code, the expected evolution $\hat{a}(\tau)$ displays a constant  \emph{critical phase} that corresponds to a decoding wave traveling towards the central positions of the graph. Further,  the critical value $\hat{a}(\tau^*)$ during such a phase does not depend on $L$, and the length of the critical phase is roughly proportional to $L$. The threshold $\epsilon^*$ is given by the maximum value of $\epsilon$ for which the critical value is exactly zero, and in this case we obtain $\epsilon^*=0.8$.\footnote{The threshold $\epsilon^*=0.8$ computed here for GPD differs from the value $\epsilon^*\approx 0.85$ obtained earlier for the terminated $A_7$ ensemble with $L=150$ (see Fig. \ref{fig:ThresRate}) due to a difference in the method of decoding the constraint codes (ML decoding in this example vs. BCJR decoding in Fig. \ref{fig:ThresRate}). If we change the component decoder from {blockwise ML to bitwise MAP (BCJR)}, the scaling law in \eqref{SL3} that predicts the finite-length performance of the SC-GLDPC code will be the same, but with different scaling parameters.}
Similar effects were first described in \cite{ou15} for (non-generalized) terminated SC-LDPC codes.

As also suggested in \cite{ou15}, using simulated decoding trajectories we should observe that $\text{Var}[a(\tau)]$ is fairly flat during the critical phase and that the covariance $\text{CoVar}[a(\tau),a(\xi)]$ decays exponentially fast with $|\tau-\xi|$, with a rate of decay that we denote by $\theta$. Fig. \ref{covar} demonstrates that this is indeed the case. In Fig. \ref{covar}(a), we show the empirical variance of the process $a(\tau)$ computed from $500$ simulated trajectories with $L=100$ and $M=2000$. In Fig. \ref{covar}(b), we show the empirical covariance $\text{CoVar}[a(\tau),a(\xi)]$ of the process obtained from the same set of simulations, where the covariance is normalized by $\text{Var}[a(\xi)]$ so that the maximum value is equal to one. Observe that an exponentially decaying function provides an accurate estimate of the normalized covariance, in which the parameter $\theta=0.87$ was obtained by a least squares fit. Based on this evidence,  the survival probability of the $a(\tau)$ process during the critical phase follows  a scaling law of the same form as the one proposed in \cite{ou15}, and thus the block error probability $P_{\text{Block}}$ can be estimated as 
\begin{align}\label{SL3}
P_{\text{Block}} \approx 1-\exp\left(-\frac{\epsilon L}{\displaystyle \frac{2\pi}{\theta}\int_{0}^{\alpha\sqrt{M}(\epsilon^*-\epsilon)}\Phi(z)\text{e}^{\frac{1}{2}z^2}\text{d}z}\right),
\end{align}
where $\Phi(z)$ is the  c.d.f. of the standard Gaussian distribution, $\mathcal{N}(0,1)$, $\epsilon L$ is the length of the critical phase, and, as in the uncoupled case, $\alpha\sqrt{M}(\epsilon^*-\epsilon)$ corresponds to the ratio of the expected number of decodable constraint nodes during the critical phase to the standard deviation of $a(\tau)$. Both $\theta$ and $\alpha$ are parameters that depend on the underlying  GLDPC block code and the edge spreading. {Given the results in Fig. \ref{27ML_simsv2} and Fig. \ref{covar},  we estimate that  $\alpha\approx5.66$. 

\begin{figure*}[t]
\begin{tabular}{cc}
 \includegraphics[width = 3.1in]{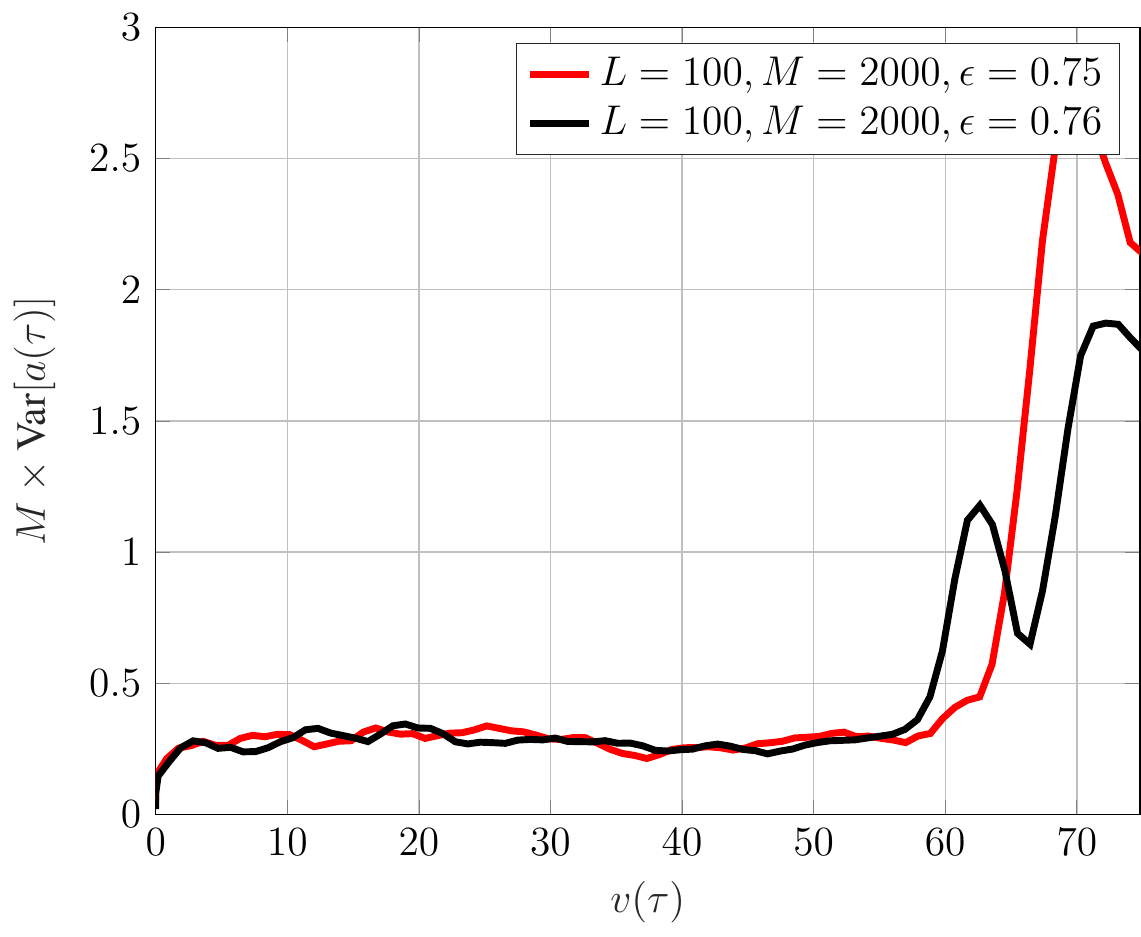} &
\includegraphics[width = 3.1in]{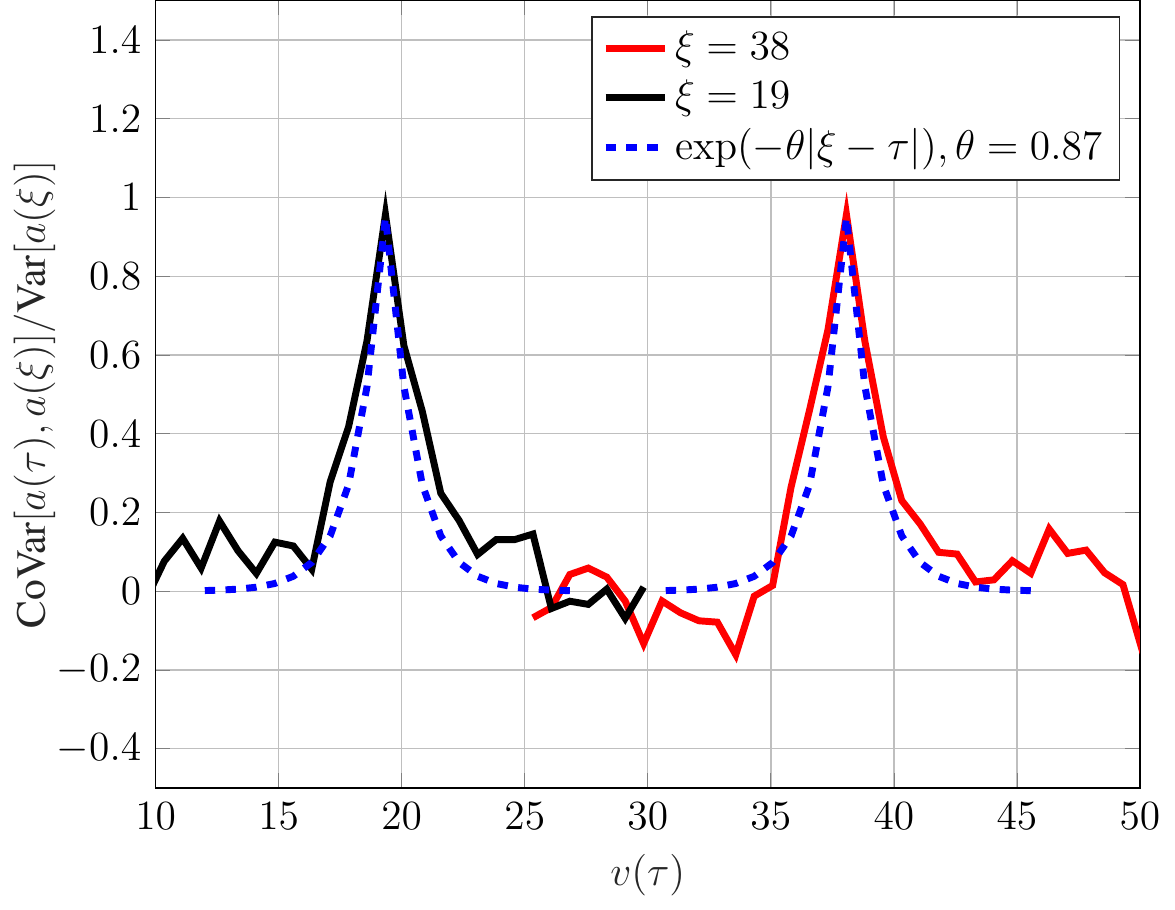} \\
(a) & (b)
\end{tabular}
\caption{Simulated trajectories of the terminated $(2,7)$-regular $A_7$ SC-GLDPC code ensemble: (a) empirical variance of the process $a(\tau)$ computed from 500 simulated  trajectories with $L=100$ and $M=2000$, and (b) empirical covariance $\text{CoVar}[a(\tau),a(\xi)]$ of the process $a(\tau)$ obtained from the same set of simulations, where the covariance is normalized by $\text{Var}[a(\xi)]$ so that the maximum value is equal to one.
}
\label{covar} 
\end{figure*}

Fig. \ref{27ML_simsv3} shows a comparison between the simulated performance (solid lines) and estimated error probability using \eqref{SL3} (dashed lines) for  the terminated $A_7$ SC-GLDPC code ensemble with $L=50$ and $(7,4)$ Hamming  constraint codes decoded with ML decoding. We again note that,  as the lifting factor $M$ increases, the performance estimate becomes very accurate.\footnote{An improved scaling law for terminated SC-LDPC codes over the BEC was recently presented in \cite{sbg19}, where the decoding process is modeled as two independent Ornstein-Uhlenbeck processes. Such an approach should also improve the performance estimate for terminated SC-GLDPC codes.} We also show the corresponding results for the uncoupled  $(2,7)$-regular GLDPC block code ensemble of Fig.~\ref{fig:27ham} with  ML-decoded $(7,4)$ Hamming constraint codes and comparable lifting factors. Besides the advantage enjoyed by the SC-GLDPC codes in decoding threshold, we see that they also exhibit better finite-length scaling behavior than GLDPC block codes, in the sense that their performance converges more quickly to the threshold.  \hfill $\Box$} 
\end{example}

%

\section{Concluding Remarks}\label{sec:conc}
Generalized LDPC (GLDPC) codes can offer significant performance improvements when compared to LDPC codes with SPC constraints at the expense of {an increase} in decoding complexity (depending on the particular constraint codes and decoders chosen), albeit with the advantage of a typically smaller number of message passing iterations. In this paper, we presented a comprehensive study of spatially coupled generalized LDPC (SC-GLDPC) codes, including both asymptotic and finite-length analyses.  Specifically, terminated SC-GLDPC code ensembles were shown to {numerically} achieve threshold saturation with near-capacity iterative decoding thresholds, thus assuring SC-GLDPC codes of having better waterfall performance than their underlying GLDPC block codes.  They were also shown to be asymptotically good and to possess large minimum distance growth rates, thus assuring them of also having excellent error floor performance.  Finally, terminated SC-GLDPC codes were shown to outperform their GLDPC block code counterparts in the finite length regime.  Based on these results, we believe SC-GLDPC codes are an attractive choice for applications requiring excellent performance throughout the entire range of decoded error rates with a limited number of decoding iterations.    


\bibliographystyle{IEEEtran}

\begin{thebibliography}{10}
\providecommand{\url}[1]{#1}
\csname url@samestyle\endcsname
\providecommand{\newblock}{\relax}
\providecommand{\bibinfo}[2]{#2}
\providecommand{\BIBentrySTDinterwordspacing}{\spaceskip=0pt\relax}
\providecommand{\BIBentryALTinterwordstretchfactor}{4}
\providecommand{\BIBentryALTinterwordspacing}{\spaceskip=\fontdimen2\font plus
\BIBentryALTinterwordstretchfactor\fontdimen3\font minus
  \fontdimen4\font\relax}
\providecommand{\BIBforeignlanguage}[2]{{%
\expandafter\ifx\csname l@#1\endcsname\relax
\typeout{** WARNING: IEEEtran.bst: No hyphenation pattern has been}%
\typeout{** loaded for the language `#1'. Using the pattern for}%
\typeout{** the default language instead.}%
\else
\language=\csname l@#1\endcsname
\fi
#2}}
\providecommand{\BIBdecl}{\relax}
\BIBdecl

\bibitem{gal63}
R.~G. Gallager, ``Low-density parity-check codes,'' Ph.D. dissertation,
  Massachusetts Institute of Technology, Cambridge, MA, 1963.

\bibitem{lmss01}
M.~G. Luby, M.~Mitzenmacher, M.~A. Shokrollahi, and D.~A. Spielman, ``Improved
  low-density parity-check codes using irregular graphs,'' \emph{IEEE
  Transactions on Information Theory}, vol.~47, no.~2, pp. 585--598, Feb. 2001.

\bibitem{tan81}
R.~M. Tanner, ``A recursive approach to low complexity codes,'' \emph{IEEE
  Transactions on Information Theory}, vol.~27, no.~5, pp. 533--547, Sept.
  1981.

\bibitem{lz99}
M.~Lentmaier and {K. Sh. Zigangirov}, ``On generalized low-density parity-check
  codes based on {Hamming} component codes,'' \emph{IEEE Communications
  Letters}, vol.~8, no.~8, pp. 248--250, Aug. 1999.

\bibitem{ypw07}
G.~Yue, L.~Ping, and X.~Wang, ``Generalized low-density parity-check codes
  based on {Hadamard} constraints,'' \emph{IEEE Transactions on Information
  Theory}, vol.~53, no.~3, pp. 1058--1079, Mar. 2007.

\bibitem{bpz99}
J.~J. Boutros, O.~Pothier, and G.~Z\'{e}mor, ``Generalized low density {Tanner}
  codes,'' in \emph{Proc. IEEE International Conference on Communications},
  Vancouver, Canada, June 1999.

\bibitem{lrc08}
G.~Liva, W.~E. Ryan, and M.~Chiani, ``Quasi-cyclic generalized {LDPC} codes
  with low error floors,'' \emph{IEEE Transactions on Communications}, vol.~56,
  no.~1, pp. 49--57, Jan. 2008.

\bibitem{mpf15}
I.~P. Mulholland, E.~Paolini, and M.~F. Flanagan, ``Design of {LDPC} code
  ensembles with fast convergence properties,'' in \emph{Proc. IEEE
  International Black Sea Conference on Communications and Networking
  (BlackSeaCom)}, May 2015, pp. 53--57.

\bibitem{fz99}
A.~{Jim\'{e}nez Felstr\"{o}m} and {K. Sh. Zigangirov}, ``Time-varying periodic
  convolutional codes with low-density parity-check matrices,'' \emph{IEEE
  Transactions on Information Theory}, vol.~45, no.~6, pp. 2181--2191, Sept.
  1999.

\bibitem{lscz10}
M.~Lentmaier, A.~Sridharan, D.~J. {Costello, Jr.}, and {K. Sh. Zigangirov},
  ``Iterative decoding threshold analysis for {LDPC} convolutional codes,''
  \emph{IEEE Transactions on Information Theory}, vol.~56, no.~10, pp.
  5274--5289, Oct. 2010.

\bibitem{kru11}
S.~Kudekar, T.~J. Richardson, and R.~L. Urbanke, ``Threshold saturation via
  spatial coupling: why convolutional {LDPC} ensembles perform so well over the
  {BEC},'' \emph{IEEE Transactions on Information Theory}, vol.~57, no.~2, pp.
  803--834, Feb. 2011.

\bibitem{kru13}
S.~Kudekar, T.~Richardson, and R.~Urbanke, ``Spatially coupled ensembles
  universally achieve capacity under belief propagation,'' \emph{IEEE
  Transactions on Information Theory}, vol.~59, no.~12, pp. 7761--7813, Dec.
  2013.

\bibitem{ftlz09}
A.~J. Felstr\"{o}m, D.~Truhachev, M.~Lentmaier, and {K. Sh. Zigangirov},
  ``Braided block codes,'' \emph{IEEE Transactions on Information Theory},
  vol.~55, no.~6, pp. 2640--2658, June 2009.

\bibitem{zlzc10}
W.~Zhang, M.~Lentmaier, {K. Sh. Zigangirov}, and D.~J. {Costello, Jr.},
  ``Braided convolutional codes: A new class of turbo-like codes,'' \emph{IEEE
  Transactions on Information Theory}, vol.~56, no.~1, pp. 316--331, Jan. 2010.

\bibitem{sfh12}
B.~Smith, A.~Farhood, A.~Hunt, F.~Kschischang, and J.~Lodge, ``Staircase codes:
  {FEC} for 100~{Gb}/s otn,'' \emph{IEEE/OSA Journal of Lightwave Technology},
  vol.~30, no.~1, pp. 110--117, 2012.

\bibitem{jpn+13}
Y.-Y. Jian, H.~D. Pfister, K.~R. Narayanan, R.~Rao, and R.~Mazahreh,
  ``Iterative hard-decision decoding of braided {BCH} codes for high-speed
  optical communication,'' in \emph{IEEE Global Communications Conference},
  Atlanta, GA, Dec. 2013, pp. 2376--2381.

\bibitem{lf10}
M.~Lentmaier and G.~Fettweis, ``On the thresholds of generalized {LDPC}
  convolutional codes based on protographs,'' in \emph{Proc. IEEE International
  Symposium on Information Theory}, Austin, TX, July 2010.

\bibitem{jian13}
Y.-Y. Jian, ``On the analysis of spatially-coupled {GLDPC} codes and the
  weighted min-sum algorithm,'' Ph.D. dissertation, Texas A\&M University,
  College Station, TX, USA, 2013.

\bibitem{mlc13}
D.~Mitchell, M.~Lentmaier, and {D. J. Costello, Jr.}, ``On the minimum distance
  of generalized spatially coupled {LDPC} codes,'' in \emph{Proc. IEEE
  International Symposium on Information Theory}, Istanbul, Turkey, 2013, pp.
  1874--1878.

\bibitem{omc15}
P.~Olmos, D.~Mitchell, and D.~Costello, ``Analyzing the finite-length
  performance of generalized {LDPC} codes,'' in \emph{Proc. IEEE International
  Symposium on Information Theory}, Hong Kong, China, July 2015, pp.
  2683--2687.

\bibitem{jpn17}
Y.~Jian, H.~D. Pfister, and K.~R. Narayanan, ``Approaching capacity at high
  rates with iterative hard-decision decoding,'' \emph{IEEE Transactions on
  Information Theory}, vol.~63, no.~9, pp. 5752--5773, Sept. 2017.

\bibitem{hpgb17}
C.~{Häger}, H.~D. {Pfister}, A.~{Graell i Amat}, and F.~{Brännström},
  ``Density evolution for deterministic generalized product codes on the binary
  erasure channel at high rates,'' \emph{IEEE Transactions on Information
  Theory}, vol.~63, no.~7, pp. 4357--4378, 2017.

\bibitem{ztk18}
L.~M. Zhang, D.~Truhachev, and F.~R. Kschischang, ``Spatially coupled
  split-component codes with iterative algebraic decoding,'' \emph{IEEE
  Transactions on Information Theory}, vol.~64, no.~1, pp. 205--224, 2018.

\bibitem{cmol18}
D.~J. Costello, D.~G.~M. Mitchell, P.~M. Olmos, and M.~Lentmaier, ``Spatially
  coupled generalized {LDPC} codes: Introduction and overview,'' in \emph{Proc.
  IEEE International Symposium on Turbo Codes and Iterative Information
  Processing}, Hong Kong, China, Dec. 2018, pp. 1--6.

\bibitem{yap18}
A.~D. Yardi, I.~Andriyanova, and C.~Poulliat, ``{EBP-GEXIT} charts over the
  binary-input {AWGN} channel for generalized and doubly-generalized {LDPC}
  codes,'' in \emph{Proc. IEEE International Symposium on Information Theory},
  Jun. 2018, pp. 496--500.

\bibitem{wf06}
Y.~Wang and M.~Fossorier, ``Doubly generalized {LDPC} codes,'' in \emph{Proc.
  IEEE International Symposium on Information Theory}, Jul. 2006, pp. 669--673.

\bibitem{tho03}
J.~Thorpe, ``Low-density parity-check ({LDPC}) codes constructed from
  protographs,'' Jet Propulsion Laboratory, Pasadena, CA, INP Progress Report
  42-154, Aug. 2003.

\bibitem{mlc15}
D.~G.~M. Mitchell, M.~Lentmaier, and {D. J. Costello, Jr.}, ``Spatially coupled
  {LDPC} codes constructed from protographs,'' \emph{IEEE Transactions on
  Information Theory}, vol.~61, no.~9, pp. 4866--4889, Sep. 2015.

\bibitem{lnf10}
M.~Lentmaier, B.~N{\"o}then, and G.~P. Fettweis, ``Density evolution analysis
  of protograph-based braided block codes on the erasure channel,'' in
  \emph{International ITG Conference on Source and Channel Coding}, Siegen,
  Germany, Jan. 2010, pp. 1--6.

\bibitem{st79}
G.~Solomon and H.~C.~A. Tilborg, ``A connection between block and convolutional
  codes,'' \emph{SIAM Journal on Applied Mathematics}, vol.~37, no.~2, pp.
  358--369, Oct. 1979.

\bibitem{mw86}
H.~H. Ma and J.~K. Wolf, ``On tail biting convolutional codes,'' \emph{IEEE
  Transactions on Communications}, vol.~34, no.~2, pp. 104--111, Feb. 1986.

\bibitem{ltf09}
M.~{Lentmaier}, M.~B.~S. {Tavares}, and G.~P. {Fettweis}, ``Exact erasure
  channel density evolution for protograph-based generalized {LDPC} codes,'' in
  \emph{Proc. IEEE International Symposium on Information Theory}, Seoul, South
  Korea, Jun. 2009, pp. 566--570.

\bibitem{mmu08}
C.~M\'{e}asson, A.~Montanari, and R.~Urbanke, ``{Maxwell} construction: The
  hidden bridge between iterative and maximum a posteriori decoding,''
  \emph{IEEE Transactions on Information Theory}, vol.~54, no.~12, pp.
  5277--5307, Dec. 2008.

\bibitem{akb04}
A.~{Ashikhmin}, G.~{Kramer}, and S.~{ten Brink}, ``Extrinsic information
  transfer functions: model and erasure channel properties,'' \emph{IEEE
  Transactions on Information Theory}, vol.~50, no.~11, pp. 2657--2673, Nov.
  2004.

\bibitem{pvc+08}
E.~{Paolini}, M.~{Varrella}, M.~{Chiani}, B.~{Matuz}, and G.~{Liva},
  ``Low-complexity {LDPC} codes with near-optimum performance over the {BEC},''
  in \emph{Proc. 4th Advanced Satellite Mobile Systems}, Aug. 2008, pp.
  274--282.

\bibitem{ips+12}
A.~R. Iyengar, M.~Papaleo, P.~H. Siegel, J.~K. Wolf, A.~{Vanelli-Coralli}, and
  G.~E. Corazza, ``Windowed decoding of protograph-based {LDPC} convolutional
  codes over erasure channels,'' \emph{IEEE Transactions on Information
  Theory}, vol.~58, no.~4, pp. 2303--2320, Apr. 2012.

\bibitem{adr11}
S.~{Abu-Surra}, D.~Divsalar, and W.~E. Ryan, ``Enumerators for protograph-based
  ensembles of {LDPC} and generalized {LDPC} codes,'' \emph{IEEE Transactions
  on Information Theory}, vol.~57, no.~2, pp. 858--886, Feb. 2011.

\bibitem{mpc13}
D.~G.~M. Mitchell, A.~E. Pusane, and D.~J. {Costello, Jr.}, ``Minimum distance
  and trapping set analysis of protograph-based {LDPC} convolutional codes,''
  \emph{IEEE Transactions on Information Theory}, vol.~59, no.~1, pp. 254--281,
  Jan. 2013.

\bibitem{tzc10}
D.~Truhachev, {K. Sh. Zigangirov}, and D.~J. {Costello, Jr.}, ``Distance bounds
  for periodically time-varying and tail-biting {LDPC} convolutional codes,''
  \emph{IEEE Transactions on Information Theory}, vol.~56, no.~9, pp.
  4301--4308, Sept. 2010.

\bibitem{amru09}
A.~Amraoui, A.~Montanari, T.~Richardson, and R.~Urbanke, ``Finite-length
  scaling for iteratively decoded {LDPC} ensembles,'' \emph{IEEE Transactions
  on Information Theory}, vol.~55, no.~2, pp. 473--498, Feb. 2009.

\bibitem{lmss01b}
M.~Luby, M.~Mitzenmacher, M.~A. Shokrollahi, and D.~A. Spielman, ``Efficient
  erasure correcting codes,'' \emph{IEEE Transactions on Information Theory},
  vol.~47, no.~2, pp. 569--584, Feb. 2001.

\bibitem{worm95}
N.~C. Wormald, ``Differential equations for random processes and random
  graphs,'' \emph{{A}nnals of {A}pplied {P}robability}, vol.~5, no.~4, pp.
  1217--1235, 1995.

\bibitem{lok19}
Y.~{Liu}, P.~M. {Olmos}, and T.~{Koch}, ``A probabilistic peeling decoder to
  efficiently analyze generalized {LDPC} codes over the {BEC},'' \emph{IEEE
  Transactions on Information Theory}, vol.~65, no.~8, pp. 4831--4853, Aug.
  2019.

\bibitem{ou15}
P.~Olmos and R.~Urbanke, ``A scaling law to predict the finite-length
  performance of spatially-coupled {LDPC} codes,'' \emph{IEEE Transactions on
  Information Theory}, vol.~61, no.~6, pp. 3164--3184, June 2015.

\bibitem{sbg19}
R.~Sokolovskii, F.~Br{\"a}nnstr{\"o}m, and A.~{Graell i Amat}, ``A refined
  scaling law for spatially coupled {LDPC} codes over the binary erasure
  channel,'' in \emph{Proc. IEEE Information Theory Workshop}, Visby, Sweden,
  Aug. 2019.

\end{thebibliography}

\end{document}